\documentclass[10pt,journal]{IEEEtran}
\usepackage{filecontents}
\usepackage{amsbsy,amsmath,amssymb,epsfig,bbm,mathrsfs,fancyhdr,fancyvrb,url}
\usepackage{graphicx}
\usepackage{float}
\usepackage{breqn}
\usepackage{chemarr}

\usepackage{amsmath}
\usepackage{amsthm}

\usepackage{enumitem}
\usepackage{mathtools}
\usepackage[utf8]{inputenc}
\usepackage{multirow}
\usepackage{amsfonts}
\usepackage{epsfig}
\usepackage{amssymb}
\usepackage{graphicx}
\usepackage{amssymb,amsmath}
\usepackage{cite}
\usepackage{color,soul}
\usepackage{amsmath}
\usepackage{color}
\usepackage{lipsum}
\usepackage{dblfloatfix}
\usepackage{bbm}
\usepackage{multicol}
\usepackage{subfigure}
\usepackage{xcolor,colortbl}
\usepackage{algcompatible}
\usepackage[noend]{algpseudocode}
\definecolor{Gray}{gray}{0.85}
\usepackage[subfigure]{tocloft}
\usepackage{epstopdf}

\newtheorem{re}{Remark}
\allowdisplaybreaks

\usepackage{amsmath, amsthm, amssymb}
\usepackage{bbding}
\usepackage[normalem]{ulem}
\usepackage{tabularx}
\usepackage{multirow}
\usepackage[normalem]{ulem}
\usepackage{array}

\usepackage{dblfloatfix}
\usepackage{multirow}
\usepackage{rotating}
\usepackage{tabularx}
\usepackage{array}
\usepackage{adjustbox,lipsum}

\usepackage{booktabs}
\usepackage{multirow}
\usepackage{tabularx}
\usepackage{array}
\usepackage{dblfloatfix}
\usepackage{hyperref}
\usepackage{blindtext}
\usepackage[normalem]{ulem} 
\usepackage{soul}
\usepackage[linesnumbered,ruled,vlined]{algorithm2e}
\SetKwInput{KwInput}{Input}
\SetKwInput{KwOutput}{Output}

\begin{document}
	
		\title{Learning based E2E Energy Efficient in Joint Radio and NFV Resource Allocation for 5G and Beyond Networks 
		 }
		\author{\IEEEauthorblockN{Narges Gholipoor,  Ali Nouruzi, Shima Salarhosseini, Mohammad Reza Javan, \IEEEmembership{Senior Member,
				IEEE}, Nader Mokari, \IEEEmembership{Senior Member, IEEE}, and Eduard A. Jorswieck,
			\IEEEmembership{Fellow, IEEE}}
				\IEEEcompsocitemizethanks{\IEEEcompsocthanksitem N. Gholipoor, Sh. Salarhosseini, A. Nouruzi and N. Mokari are with the Department of ECE,
		Tarbiat Modares University, Tehran, Iran (email: \{Gholipoor.narges, shima.salarhosseini and nader.mokari\}@modares.ac.ir). 
	Mohammad R. Javan is with the Department of Electrical and Robotics Engineering, Shahrood University of Technology,
	Shahrood, Iran (javan@shahroodut.ac.ir).
	Eduard A. Jorswieck is with  TU Braunschweig, Department of Information Theory and Communication Systems, Braunschweig, Germany (jorswieck@ifn.ing.tu-bs.de).
}}
	\maketitle
	\begin{abstract}
In this paper, we propose a joint radio and core resource allocation framework for NFV-enabled networks. In the proposed system model, the goal is to maximize energy efficiency (EE), by guaranteeing end-to-end (E2E)  quality of service (QoS) for different service types. To this end, we formulate an optimization problem in which power and spectrum resources are allocated in the radio part. In the core part, the chaining, placement, and scheduling of functions are performed to ensure the QoS of all users.
This joint optimization problem is modeled as a Markov decision process (MDP), considering time-varying characteristics of the available resources and wireless channels. A soft actor-critic deep reinforcement learning (SAC-DRL) algorithm based on the maximum entropy framework is subsequently utilized to solve the above MDP.
Numerical results reveal that the proposed joint approach based on SAC-DRL algorithm  could significantly reduce energy consumption compared to the case in which R-RA and NFV-RA problems are optimized separately. 

\end{abstract}
\maketitle
\section{{introduction}}
\IEEEPARstart{T}{o} support the exponential growth of traffic and various services, communication service providers (CSPs) need to redesign their infrastructure or move toward a flexible programmable infrastructure \cite{8320765}. On the other hand, CSPs seek to increase capacity demands and upgrade their network in the shortest possible time. Network function virtualization (NFV) and softwarization are the key technologies that can meet the requirements of increasing the exponential traffic, various quality of services (QoS) requirements intended in the next generations of cellular networks, i.e., fifth-generation  (5G) wireless cellular network. \cite{8125672}.
  NFV is emerged as a critical technology to reduce the network CAPEX and OPEX, and time to market by virtualizing all the appliances such as servers, routers, storage, and switches \cite{mijumbi2015design, mijumbi2016network,herrera2016resource, 8675284,riera2014virtual}.
 NFV technology allows network functions to be run virtually on network servers as virtual network functions (VNFs), e.g.,  firewall, deep packet inspection, transcoding, and load balancing \cite{8125672}. In this paper, we assume that  VNF and network function (NF) are the same  \cite{alliance2016description,7926921,7243304}.
 

 \subsection{Background to NFV and Radio Resource Allocation}
One of the critical challenges in the NFV-based network is resource allocation (RA) \cite{herrera2016resource,7243304,riera2014virtual}.   NFV resource allocation (NFV-RA) consists of three phases: 1) The first phase is  the VNF service function chaining (VNF-SFC), in which the chain and the connection of VNFs are determined \cite{7945848}, 2) The second phase is VNF Placement, in which the VNFs are mapped to the servers/virtual machines (VMs), and 3) The third phase is VNF scheduling in which the running time for each VNF is determined \cite{7243304}. Each of the mentioned phases has a significant impact on the network performance, e.g., total cost and also QoS of users. Therefore, optimizing all these phases together can have a significant impact on reducing total costs and improving QoS of users, e.g., end to end (E2E) delay.
Besides, in the radio part, we are faced to limited power resources and spectrum scarcity; thus, it is necessary to efficiently allocate  radio resources, i.e., radio resource allocation (R-RA),  to provide a variety of services while ensuring the QoS of each user. Therefore, to provide an E2E service, the NFV-RA and R-RA should be considered jointly \cite{hossain20155g, 7143328}.


Network services have various requirements such as delay and data rate. Since delay is one of the requirements that affect all parts of the network, to provide services that require a specific delay, it is necessary to consider all parts of the network together. Also, due to the fact that different parts of the network interact with each other, it is necessary to consider all parts of the network together to optimize the total energy consumption.
Machine learning (ML) are new approaches to solve optimization problems in all domains, e.g., network domain, that have recently received a great deal of attention \cite{8715830,8932445,8354944}.  ML, due to its data-driven nature, automatically learns the network and communication environment and dynamically adapts protocols without human intervention \cite{8715830,8932445,8354944}.
\subsection{{\textcolor{black}{ Related Works}}}
 
In this section, we review the related works. For this purpose, we classify the related works into two main categories: NFV-based works and learning methods for RA.

\subsubsection{NFV-RA}
As mentioned above, the NFV-RA problems are divided into three categories: 1) SFC problems, 2) VNFs Placement problems, and 3) VNFs scheduling problems. Each of these categories is discussed in the following.
	\paragraph{VNF Chaining}
In the NFV-based network, the flow should be passed through a sequence of middleboxes in a particular order called SFC to provide a service \cite{xie2016service}. A resource allocation architecture which enables energy-aware SFC
		for a software-defined network (SDN) based is proposed in \cite{8480442} by considering constraints on delay, link utilization, and server utilization. To solve this problem, a heuristic algorithm is proposed.             In \cite{7417401}, a VNF chaining problem is addressed to minimize bandwidth utilization while computing results within reasonable runtime. To solve the problem, a heuristic algorithm is proposed. In \cite{8170213}, a  VNF chaining problem is formulated to minimize total CAPEX and OPEX costs. In \cite{liu2017dynamic}, a VNF chaining problem, while considering the trade-off between resource consumption and operational overhead, is studied. An integer linear programming (ILP) model is utilized to solve this problem. In \cite{81702199999},  a VNF Selection and chaining problem in SDN/NFV-enabled networks by considering minimum E2E delay is investigated.
\paragraph{VNF Placement}
In \cite{riggio2016scheduling}, the VNF placement problem is formulated in which the computational resources such as processing and storage capacity of CPUs and the capacity of virtual links are considered. This problem is solved by a heuristic algorithm. 	In \cite{cohen2015near}, a VNF placement problem is proposed to minimize the total cost by considering the limited space for allocating functions in each server. A linear relaxation approximation is used to solve this problem.
An algorithm for VNF placement and CPU assignment to VNFs is proposed in \cite{8611305}, in which the proposed problem is solved by decoupled solution strategy leveraging on sequential decision making. A problem of placement and chaining is studied in \cite{8255993}. In this problem, nodes and links  latency and capacities are considered. In \cite{cao2017vnf}, the VNF placement optimization problem is formulated for achieving lower bandwidth consumption and lower maximum link utilization. A genetic algorithm is proposed to solve this problem. In \cite{8316876}, a  VNFs placement and chaining problem is formulated to minimize the total cost, including the cost of deploying VNF instances, the cost of using surrogate servers, and the cost of communication. Moreover,  in that work, the communication delays and processing delays of VNFs are considered. To solve the proposed algorithm, a heuristic algorithm is presented. A VNF placement cost minimization problem, taking the network stability into account, is formulated in \cite{8424410}. The proposed problem is decomposed into an SFC and VNF mapping problem, which are solved by a genetic algorithm-based heuristic.
In \cite{8281644},  VNF placement in the cloud data center to minimize the number of activated physical machines and consider time-varying workloads of physical machines is studied. The proposed problem is formulated as an integer linear programming (ILP) model, which is solved by a greedy-based algorithm.
 A Fully decentralized approach for online placement and optimization of VMs for NFV-based network is proposed in \cite {8501940}.
 	 A VNF placement to optimize operational and traffic cost is formulated in \cite{7859379}. To solve that problem,  a sampling-based Markov approximation (MA) approach is applied which is a combination of Markov and matching algorithm. {The} joint VNF placement and admission control to maximize the network provider revenue in terms of bandwidth and capacity are investigated in \cite{nejad2018vspace}. The relaxation, reformulation, and successive convex approximation (SCA) methods are employed to solve this problem. In \cite{9000731}, a cross-layer resource optimization model and solution for wireless-enabled SFC is proposed to optimize VNF placement and route path by considering E2E downlink latency, including both wireless and wired delay. 
	\paragraph{VNF Scheduling}
	In \cite{mijumbi2015design}, a VNF scheduling and placement algorithm is proposed in which the available buffer capacity and processing time of the VNFs are considered. A greedy-based method is used to solve the proposed algorithm. In \cite{riera2014virtual}, a scheduling problem for SDN-based system is proposed, which is solved with a two-stage approach. In \cite{7490404}, a  joint VNF scheduling and traffic steering problem is formulated in which VNF transmission and processing delays are considered. To solve this problem, a genetic algorithm-based method is utilized. A matching-based algorithm for solving the VNF scheduling problem is presented in \cite{game-theroy}. A deadline-aware VNF scheduling problem for ultra-low latency services is presented in \cite{8256017} to maximize the number of admitted services. The proposed problem is solved by utilizing a tabu search-based heuristic method.  In \cite{kim2018performance},  an RA algorithm for VNFs placement and scheduling is studied.

\subsubsection{Learning Methods for RA}
In \cite{8951149}, a machine learning-enabled architecture is investigated to provide the demands of advanced vehicular Internet infrastructures. In \cite{8932445}, a VNF placement problem in SDN/NFV-enabled networks is formulated. To solve the proposed problem, a deep reinforcement learning (RL) Algorithm is proposed. A multi-objective resource allocation problem for NFV orchestration (NFVO), as a Markov decision process, is formulated in \cite{9128963} in which A Q-learning based algorithm is proposed to solve this multiple objectives problem. In \cite{8855889}, a problem of scheduling VNFs is studied to minimize the overall completion time of all services by considering E2E delay requirements. To solve the problem with high efficiency and high accuracy, the authors utilized the RL algorithm to find the optimal scheduling policy. In \cite{8901169}, a novel method based on RL for performing dynamic SFC resource allocation in NFV-SDN enabled network is proposed. To design intelligent and efficient VNF selection and chaining for service function chaining requests, a deep learning-based two-phase algorithm is introduced in \cite{81702199999}.

 In all of the above works, the authors consider only one server hosts several VMs  or several servers each of which hosts only one VM. The critical point is that several servers that host several VMs are  not considered simultaneously in any of the above works. However, in real systems, each server can process multiple VMs simultaneously. 
 As a result, considering multiple VMS on each server increases the complexity of the system model and requires new solutions. Most importantly, to provide some of the 5G and beyond (5G+) services, such as ultra-low latency services, the network should be considered from an E2E perspective which is not considered in all previous works. Furthermore, due to the resource limitation and 5G+  services requirements such as E2E delay, it is necessary to consider and optimize the NFV and radio parts jointly.  Moreover, the main drawback of the above works is that the new solution methods, such as ML methods, are not applied to solve the optimization problem in the NFV-RA area.

\subsection{\textbf{Our Contributions}}
 The main contributions of this paper can be summarized as follows:
\begin{itemize}
	\item In this paper, a joint NFV-RA and R-RA optimization problem from the E2E perspective is proposed, in which QoS of the requested services are satisfied.
			In our proposed problem, our aim is to minimize the energy per service by considering the limitation of  radio resources, i.e., power and spectrum, as well as the computational and storage capacities of core servers. Moreover, we consider multiple VMs per server in this work, which is a more realistic model and leads to increase in the complexity of formulation and, thus, the problem solution.
	\item We introduce a new approach for  VNF chaining, placement, and scheduling by considering the network service delay in the closed-form expression with mathematical representation.

\item Since the VNF chaining, placement, scheduling, and E2E-QoS for different services and radio resources satisfy Markov property, we model the optimization problem as a Markov decision process (MDP).  The proposed MDP problem is subsequently solved by utilizing a state-of-the-art off-policy deep reinforcement learning (DRL) algorithm, namely soft actor-critic \cite{SOFT}, which is based on the maximum entropy framework.
	
\item A Python-based simulator is developed in the simulation to implement the proposed algorithm and other baseline algorithms.
 
	\item \textcolor{black}{We provide numerical results for the performance evaluation of joint R-RA and NFV-RA algorithms for different network configurations. Our simulation results reveal joint R-RA and NFV-RA outperforms conventional ones by approximately  100 \% by increasing computational complexity.}
\end{itemize}
\subsection{\textbf{Paper Organization}}
The rest of the paper is outlined as follows.  In Section \ref{systemmodel},  the system model and problem formulation {are} explained. 
The proposed solution is presented in Section \ref{solutions}.  In Section \ref{Complexity-Convergance}, the computational complexity and convergence of our solution are discussed. The simulation results are presented in Section \ref{simulations}. Finally, in Section \ref{Conclusion}, the conclusion remarks is inferred.

\textcolor{black}{\textbf{Symbol Notations:} Vector and matrices are indicated by bold
lower-case and upper-case characters, respectively. 
 $\mathcal{A}$ denotes set $\{1,\dots,A\}$, $a(i)$ is {the} $i$-th element of set $\mathcal{A}$,  and $\Bbb{R}^{n}$ is the set of $n$ dimension real numbers.  Moreover, $U_d[a,~b]$ denotes the uniform distribution in interval $a$ and $b$ and $ |.| $ indicates absolute value.} 
\section{{system Model and problem formulation}} \label{systemmodel}
\begin{figure*}
	\centering
	\centerline{\includegraphics[width=0.7\textwidth]{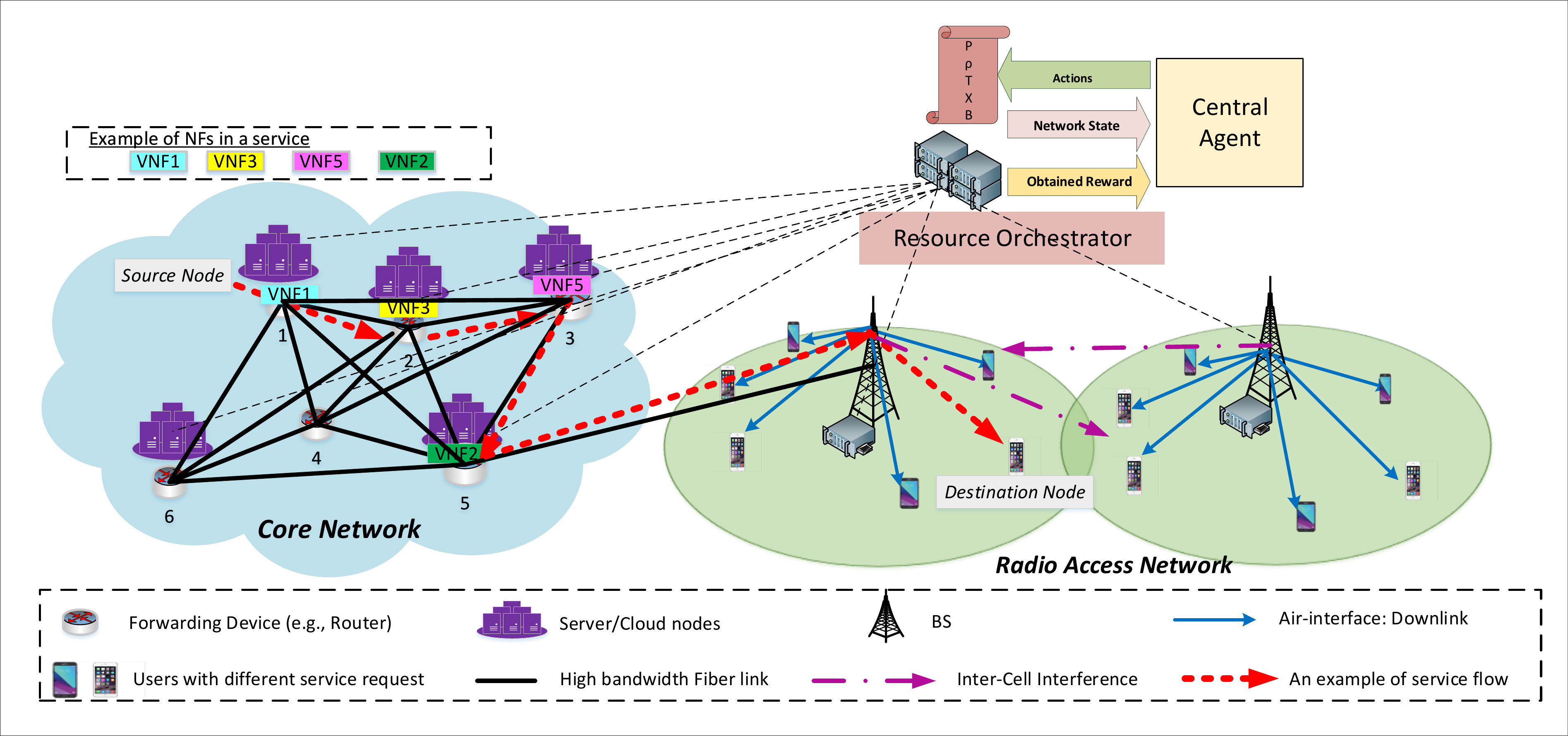}}
	\vspace{-1em}
	\caption{Considered system model  with an example E2E flow  for a user is denoted by red dashed line. In this system model, the resource orchestrator derives all of the optimization variables. 
  }
	\label{pic}
\end{figure*}

As shown in Fig. \ref{pic},  we consider an E2E network consisting of a multi-cell radio access part and an NFV-based core part with several servers/nodes in this system model. Therefore, we describe the system model from both aspects, i.e., radio part and core part,  in the following.

\begin{re}
It is worth noting that the wireless channel state information (CSI), i.e., radio access network (RAN) parameters,  changes faster than the parameters of the NFV, i.e., SFC. In this paper, we assume that the parameters of the radio and NFV parts are fixed in each optimization problem similar to the existing works \cite{9000731, 7949048}.
\end{re}
		 
		
\subsection{\textbf{Radio Access Network Description}}
We consider a multi-cell network with  a set of  $\mathcal{J}=\{1,\dots,J\}$ BSs which are connected to the core part. Moreover, we consider a set of $\mathcal{U}=\{1,\dots,U\}$	users
	and a set     $\mathcal{K}=\{1,\dots,K\}$ subcarriers with bandwidth $B$. We define the subcarrier assignment variable $\rho_{u,j}^{k}$ with $\rho_{u,j}^{k}=1$ if subcarrier $k$ is allocated to user $u$ at BS $j$ and otherwise $\rho_{u,j}^{k}=0$.   We assume orthogonal frequency division multiple access (OFDMA) as the transmission technology in which each subcarrier is assigned at most to one user in each BS. Thus, the following constraint is introduced:
\begin{equation} \label{subcons}
\text{C1:~}\sum_{u \in \mathcal{U}} \rho_{u,j}^{k} \le 1, \forall k\in\mathcal{K}, j \in \mathcal{J}.
\end{equation}
Let $h_{u,j}^{k}$ be the channel coefficient between user $u$ and 
BS $j$ on subcarrier $k$, $~p_{u,j}^{k}$ be the transmit power from  BS $j$ to user $u$ on subcarrier $k$, $\sigma_{u,j}^{k}$ be the power of additive white Gaussian noise (AWGN)
for user $u$ at BS $j$ on subcarrier $k$, and $I_{u,j}^{k}$ the is the inter-cell interference on user $u$ at BS $j$ over subcarrier $k$. 
Thus, the received signal to interference and noise ratio (SINR) of user $u$ at BS $j$ on
subcarrier $k$ is $ \gamma_{u,j}^{k}=\frac{ p_{u,j}^{k} h_{u,j}^{k}}{\sigma_{u,j}^{k}+I_{u,j}^{k}}$, and the achievable data rate (in bits per second/Hz) of user $u$ at BS $j$ on subcarrier $k$ is given by
\begin{align}
r_{u,j}^{k}= \rho_{u,j}^{k}\log(1+\gamma_{u,j}^{k}), \forall u \in \mathcal{U}, k \in \mathcal{K}, j \in \mathcal{J}.\label{Rate_for}
\end{align}

\textcolor{black}{Hence, the total achievable rate of user $u$ at BS $j$ is given by
$R_{u,j}=\sum_{k \in \mathcal{K}}r_{u,j}^{k},\,\forall u\in\mathcal{U}, j \in \mathcal{J}$.
The transmit power limitation of BS $j$ is
\begin{equation}
\text{C2:~}\sum_{k\in \mathcal{K}}  \sum_{u\in \mathcal{U}} \rho_{u,j}^{k}  p_{u,j}^{k}  \le P_j^{\max}, \forall j \in \mathcal{J},
\end{equation}
 where $P_j^{\max}$ is the maximum transmit power of BS $j$.}
\subsection{{NFV Environment Description}}
Here, we illustrate how the generated traffic of  each user is handled in the network by performing different NFs in the requested user's NS\footnote{Defined by European Telecommunications Standards Institute (ETSI) as the composition of Network Function(s) and/or Network Service(s), defined by its functional and
	behavioral specification \cite{ETSIG003}.} on the different servers/physical nodes by leveraging NFV\footnote{Standardized by  ETSI organization for 5G and beyond \cite{etsi2013network}.}.  
In this regard, we consider NFV-RA that consists of a new approach for the placement and scheduling phases. In the placement phase, we map each NF on the server that is capable to run that NF. Moreover, in the scheduling phase, all NFs are scheduled on each server. Note that we do not consider mapping virtual links on the physical links and leave it as an interesting future work as \cite{mijumbi2015design, yoon2016nfv}. Besides, similar to \cite{9076109, 8423711,7387398}, we assume that the core network is full-mesh and there is a direct path between any two nodes in the network.

	\begin{table}[t]	
		\renewcommand{\arraystretch}{1.5}
		\centering
		\caption{Network parameters and notations}
		\begin{adjustbox}{width=.48\textwidth,center}	
			\begin{tabular}{| c| l| }	
				\hline
				\textbf{Notation}& \textbf{Definition}\\\hline
				$\mathcal{J}/J/j$&Set/number/index of BSs\\ \hline
				$\mathcal{U}/U/u$&Set/number/index of users\\ \hline
				$\mathcal{K}/K/k$&Set/number/index of subcarriers with bandwidth $B$\\ \hline
				$\mathcal{N}/N/n$ &Set/number/index of  servers/physical nodes\\ \hline
				$\mathcal{V}/V/v$ &Set/number/index of  VMs\\ \hline
				$\mathcal{F}/F/f$ &Set/number/index of  NFs\\ \hline
				$\mathcal{S}/S/s$ &Set/number/index of  services\\ \hline
				$\Omega_{s}$ & Set of NFs which constructs service $s$ \\ \hline
				$D^{\text{max}}_{s}$ & Delay constraint for each packet of service $s$ \\ \hline
				$R^{\min}_{s}$ &  Minimum required data rate of service $s$\\ \hline
				$P_j^{\max}$&Maximum transmit power of the BS	$j$\\\hline
				${\rho_{u,j}^k}$ &Assignment of subcarrier $k$ to user $u$ at BS $j$ \\\hline
				${p_{u,j}^k}$ &Transmit power of user $u$ from BS $j$ on subcarrier $k$ \\\hline
				${h_{u,j}^k}$ &Channel coefficient between user $u$ and the BS $j$ on subcarrier $k$\\\hline
				$\gamma_{u,j}^{k}$&SINR of user $u$ at BS $j$ on subcarrier $k$\\\hline
				$r_{u,j}^{k}$ &Achieved rate of user $u$ at BS $j$ on subcarrier $k$\\\hline	
				$y_{u,j}$ & Packet size of the requested service of user $u$ at BS $j$ \\\hline
				$q_{v,n}^{f_{m}^{s}}$, $\psi_{v,n}^{f_{m}^{s}}$& Processing and storage demand of NF $f_{m}$ in service $s$ on VM\\&  $v$ on server $n$, respectively\\\hline
				$\tau_{u,j,n}^{f_{m}^{s},v}$& Processing delay of NF $f_{m}$ on VM $v$ on server $n$ for service $s$\\\hline
				$L_n, \Upsilon_n$ & Processing and storage capacity  of server $n$, respectively\\\hline
				$\tilde L_v^n, \tilde \Upsilon_v^n$ & Required Processing and storage capacity  for running VM $v$ on \\&server $n$, respectively\\\hline
				$e_v^{f_m^s}$ & VM $v$ hosts NF $f_m^s$\\\hline
				$\chi_{u,j}^s$ & User $u$ at BS $j$ requests service $s$ \\\hline
				$\beta_{u,j,n}^{f_{m}^{s},v}$ & Server mapping between  NF $f_{m}^{s}$ of service $s$ for user $u$, and \\&VM $v$ hosts on node $n$\\\hline
				$q_{v,n}^{f_{m}^{s}}$ & Required processing capacity to run NF ${f_{m}^{s}}$  on VM $v$ over the \\&assigned server $n$ \\\hline
				$\psi_{v,n}^{f_{m}^{s}}$ & Required storage capacity to run NF ${f_{m}^{s}}$  on VM $v$ over the \\&assigned server $n$ \\\hline
				$t_{u,j,n}^{f_{m}^{s},v}$ & Start time of running NF $f_{m}^s$ of the requested service $s$ for user\\& $u$ at BS $j$ on server $n$ in VM $v$\\ \hline
				$x_{u,j,n,u',j',n'}^{f_{m}^{s},v,f_{m'}^{s'},v'}$& Ordering indicator between NF $f_{m}^{s}$  of service $s$ for user $u$ and\\& NF  $f_{m'}^{s'}$ of service $s'$ for user $u'$\\ \hline
			\end{tabular}
		\end{adjustbox}\label{Table_parameters}
	\end{table}	
	
We consider $S$ NS 
types 
whose set is $\mathcal{S}=\{1,2, ..., S\}$ and $M$ NFs whose set is  $\mathcal{F} = \{f_{m}~\big|~m=1,\dots,M\}$. 
Hereinafter, we use the term service instead of NS.
The considered parameters of the paper are stated in Table \ref{Table_parameters}.
Each service $s$ is defined by the tuple ${\Big(}\textit{source node},\Omega_{s},D^{\text{max}}_{s},R^{\min}_{s}{\Big)}$\footnote{\textcolor{black}{We assume that the input node is specified for each service, and besides, the output node of the path is the user requesting the service.}} where $\Omega_{s}$ is the set of NFs which constructs service $s$ defined by $\Omega_{s}=\left\{f_{m}^{s}\right\},\,m\in\{1,\dots,M\}$, $D^{\text{max}}_{s}$ is the delay constraint for each packet of service $s$, and $R^{\min}_{s}$ is the minimum required data rate of service $s$. Here, we define a known binary variable $\chi_{u,j}^s \in \{0,1\}$ that indicates user $u$ at BS $j$ requests service $s$.  Based on the data rate requirement of each service, we have:
\begin{align} 
\text{C3:~}R_{u,j} \ge \sum_{s \in \mathcal{S}}\chi_{u,j}^s R^{\min}_{s},\, \forall u\in\mathcal{U}.\nonumber
\end{align}

We consider a set of physical node/servers denoted by  $\mathcal{N} = \{1, ..., N\}$ in the network each of which has a limited amount of computing and storage resources. The CPU and storage resources of server $n$ is defined by the tuple ${\Big(} L_n, \Upsilon_n{\Big)}$ where $L_n$ is processing capacity (CPU cycle per unit time) and  $\Upsilon_n$ is storage capacity of server $n$ (number of packets per unit time). Moreover, we consider a set of VMs indicated by $\mathcal{V} = \{1, ...,V\}$ running on top of physical nodes and can host one or more NFs. Besides, VM $v$ needs a processing resource ($\tilde L_v^n$) and storage resource ($\tilde \Upsilon_v^n$) to run on server $n$.
Furthermore, we define a known binary variable $e_v^{f_m^s}$  with $e_v^{f_m^s}=1$ if VM $v$ can host NF $f_m^s$ and otherwise $e_v^{f_m^s}=0$.  
We assume that each VM can process at most one function at each time \cite{mijumbi2015design}, but it can process any NF \cite{mijumbi2015design}, if capable to run it based on variable $e_v^{f_m}$. 



To embed and schedule NFs, we first need to determine which server hosts which VM. Then, it must be determined how the NFs are to be scheduled in the VMs and servers.
We introduce a binary variable $\beta_{u,j,n}^{f_{m}^{s},v}$ (i.e., VNF-placement variable) which denotes that  NF $f_{m}^{s}$ of service $s$ for user $u$ at BS $j$ is executed at node $n$ in VM $v$, and is defined as
\begin{equation} \nonumber
\begin{split}
&\beta_{u,j,n}^{f_{m}^{s},v}= \begin{cases}
1, & \text{NF~ $f_{m}^{s}$~of service~$s$~for~user~$u$~at BS~$j$~ is}\\ & \text{executed~at server~$n$ in VM $v$}.\\
0, & \text{Otherwise}.
\end{cases}
\end{split} 
\end{equation}
Each NF of each service is performed completely at only one VM and server at a time \cite{7490404}. Therefore, we have
\begin{align} 
\text{C4:~}\sum_{n \in \mathcal{N}}\sum_{v \in \mathcal{V}}\beta_{u,j,n}^{f_{m}^{s},v}e_v^{f_m^s} \le 1, \forall u \in \mathcal{U},  j \in \mathcal{J}, f_{m}^{s} \in \Omega_{s}, s\in\mathcal{S}.\nonumber
\end{align}	

Moreover, we assume  that  NF ${f_{m}^{s}}$ needs a specific number of CPU cycles per bit  to run on VM $v$ over the assigned server $n$, i.e., $q_{v,n}^{f_{m}^{s}}$. From the physical resource perspective, we assume that each server $n$ can provide at most $L_n$ CPU cycles per unit time. 
Therefore, by considering total required CPU cycles, we have the following constraint:
\begin{align} 
\text{C5:~}\sum_{j \in \mathcal{J}}
\sum_{u \in \mathcal{U}}
\sum_{s\in\mathcal{S}}\sum_{v\in\mathcal{V}}\sum_{f_{m}^{s}\in\Omega_{s}}
\beta_{u,j,n}^{f_{m}^{s},v} e_v^{f_m^s}\big({y}_{u,j}q_{v,n}^{f_{m}^{s}}+\tilde L_v^n\big) \le L_n ,\nonumber\\ \forall n \in \mathcal{N},\nonumber 
\end{align}
where $y_{u,j}$ is the packet size of the service of user $u$ at BS $j$. Here, we assume that the packet size is equal to the number of bits generated in a unit time \cite{mijumbi2015design,riera2014virtual,7490404}.
Hence, the elapsed time of each NF $f_{m}^{s}$ for each bit on VM $v$ over server $n \in \mathcal{N} $ 
	is equal to $ \frac{1}{q_{v,n}^{f_{m}^{s}}},\forall n\in\mathcal{N}, f_{m}^{s} \in \Omega_{s}$. 
Therefore, the total processing delay of running  NF $f_{m}^{s}$ on VM $v$ at server $ n $ for each packet with packet size 
${y}_{u,j}$ of user $u$ at BS $j$ is obtained as 
\begin{align}\label{Final_Del}
\tau_{u,j,n}^{f_{m}^{s},v}=	\frac{{y}_{u,j}}{q_{v,n}^{f_{m}^{s}}},\forall n\in\mathcal{N}, f_{m}^{s} \in \Omega_{s}.
\end{align}


Additionally, we assume that each NF needs specific storage size, i.e., $\psi_{v,n}^{f_{m}^{s}}$, when it is running on VM $v$ on the server $n$. Moreover, each packet  consumes  $y_{u,j}$ buffer capacity, when it is waiting for running a NF on an assigned server.  Hence, from the storage and buffer resource perspective, we consider that each server has a limited buffer and storage size, i.e., $\Upsilon_n$, which leads to the following constraint:
\begin{align}
	\text{C6:~}
\sum_{u\in\mathcal{U}}\sum_{\substack{{s\in\mathcal{S}}\\f_{m}^{s}\in\Omega_{s}}}\sum_{j \in\mathcal{J}}\sum_{v\in\mathcal{V}}
	 \beta_{u,j,n}^{f_{m}^{s},v}e_v^{f_m^s} \Big((\psi_{v,n}^{f_{m}^{s}}  +
	{y}_{u,j} )+\tilde \Upsilon_v^n\Big)\le \Upsilon_n ,&\nonumber\\ \forall n \in \mathcal{N}.& \nonumber
\end{align}
\subsection{\textbf{Latency Model}}
In NFV-RA, our main aim is to guarantee the service requirement, which includes maximum tolerable delay  for each packet with size ${y}_{u,j}$ of the requested services while minimizing the energy consumption of physical nodes \cite{mijumbi2015design,riera2014virtual,7490404}.  
The total delay that we consider in our system model results from executing NFs, waiting time, and transmission time between nodes. In the following, we calculate the total delay resulting from scheduling.


Each NF should wait until its preceding function is processed before its processing can commence.
The processing of each packet of service $s$ ends when its last function is processed. Therefore, the total processing time is  the summation of the processing times of the NFs at the various VMs on servers. 
	For scheduling of each NF on a server,  we need to determine the start time of it. Therefore, we define $t_{u,j,n}^{f_{m}^{s},v}$ which is the start time of running NF $f_{m}^s$ of the requested service $s$ for user $u$ at BS $j$ on server $n$ in VM $v$. Furthermore, we introduce a new variable $x_{u,j,n,u',j',n'}^{f_{m}^{s},v,f_{m'}^{s'},v'},$ 
in which, if NF $f_{m}^{s}$ of user $u$ at BS $j$ is running on VM $v$ at server $n$ after NF $f_{m'}^{s'}$ of user $u'$ at BS $j'$ is running on VM $v'$ at server $n'$, its value is $1$, otherwise is $0$. 
By these definitions, the starting time of each NF can be obtained in \eqref{schule}.  The first part of this equation determines the priority of performing the NFs of the requested service by a specific user. The second part determines the priority of running each NF of user $u$ in relation to each NF of user $u’$. Thus, the execution time and location of each NF are determined.  In the following, we  clarify this equation with an example.   
\begin{figure*}
		\begin{align}
		&\text{C7:~} t_{u,j,n}^{f_{m}^{s},v}\beta_{u,j,n}^{f_{m}^{s},v}e_v^{f_m^s}\ge  \max \Bigg\{ \mathop {\max }\limits_{\scriptstyle\forall f_{m'}^{s'}\in\Omega_{s'},
			\scriptstyle u'\in\mathcal{U}} \left\{ {x_{u,j,n,u',j',n}^{f_m^{s},v,f_{m'}^{s'},v}\beta _{u',j',n}^{f_{m'}^{s'},v}e_v^{f_{m'}^{s'}}(t_{u',j',n}^{f_{m'}^{s'},v} + \tau _{u',j',n}^{f_{m'}^{s'},v})} \right\},
			\nonumber	\\ &
		\mathop {\max }\limits_{\scriptstyle\forall f_{m''}^s\in \{{\Omega _s}-f_{m}^{s}\},
			\scriptstyle  n'\in \{\mathcal{N}-n\} } \left\{ {x_{u,j,n,u,j,n'}^{f_{m}^{s},v,f_{m''}^{s},v'}\beta _{u,j,n'}^{f_{m''}^s,v'}e_{v'}^{f_{m''}^s}(t_{u,j,n'}^{f_{m''}^{s},v'} + \tau _{u,j,n'}^{f_{m''}^{s},v'})} \right\}\Bigg\},
		\nonumber\\&
		\forall f_m^s \in {\Omega _s}, f_{m'}^{s'}\in{\Omega_{s'}},{\mkern 1mu}  \forall s, s'\in\mathcal{S}, \forall n \in \mathcal{N},\, \forall u,u'\in \mathcal{U}, \forall v , v' \mathcal{V}, \forall j,j'\in \mathcal{J}. \label{schule}
		\end{align}
			\hrule
\end{figure*}

\begin{figure*}[t]
		\centering
		\centerline{\includegraphics[width=0.75\textwidth]{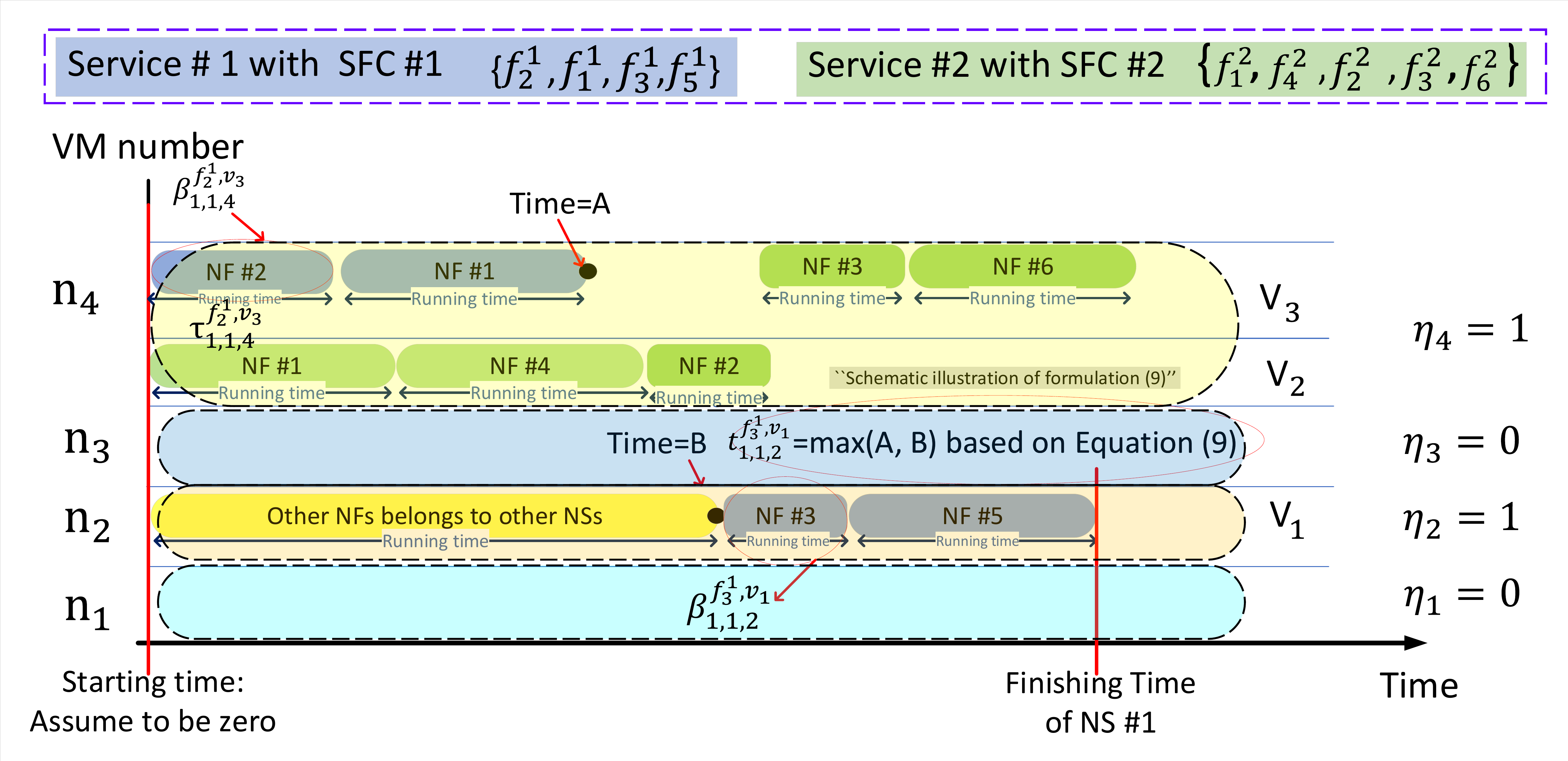}}
		\vspace{-1em}
		\caption{\textcolor{black}{Schematic illustration of the proposed scheduling and formulation of  \eqref{schule}.}}
		\label{Scheduling}
	\end{figure*}

		To demonstrate  how we formulate the scheduling of NFs, the proposed scheduling policy is illustrated  in Fig. \ref{Scheduling}.  
This figure is the state of the network  assuming two services with several NFs, four servers,  and three VMs.
 The processing time of each NF, $\tau_{u,j,n}^{f_{m}^{s},v}$ is obtained by \eqref{Final_Del}, i.e., $\tau_{1,1,4}^{f_{2}^{1},v3}=\frac{y_{1,1}}{q_{v3,4}^{f_{2}^{1}}}$. Each NF has a start time (denoted by $t_{u,j,n}^{f_{m}^{s},v}$) to run on an assigned VM and elapsed processing time (denoted by $\tau_{u,j,n}^{f_{m}^{s},v}$) and 
is completed by the time  given by $t_{u,j,n}^{f_{m}^{s},v}+\tau_{u,j,n}^{f_{m}^{s},v}$ on VM $v$ at server $n$. 
To make \eqref{schule} clearer, suppose user  $1$ at BS $1$ requests service $1$, and user $2$ at BS $2$ requests service $2$.
As can be seen, for user 1 $f_1^1$ is executed after $f_2^1$ at VM $3$ on server $4$. Thus, we have 
$x_{1,1,4,1,1,4}^{f_{1}^{1},v3,f_{2}^{1},v3}=1$ and $x^{1,1,4,1,1,4}_{f_{1}^{1},v3,f_{2}^{1},v3}=0$.
Moreover, as shown in this figure, $f_3^1$ of user $1$ at BS $1$ is executed at VM $1$ on server $2$ after $f_1^2$ of user $2$ at BS $2$ is executed at VM $2$ on server $4$. Thus, we have $x_{1,1,2,2,2,4}^{f_{3}^{1},v1,f_{1}^{2},v2}=1$ and $x^{1,1,2,2,2,4}_{f_{3}^{1},v1,f_{1}^{2},v2}=0$.
As can be seen, 
servers $3$ and $1$ are off, since based on our aim and solution algorithm, two servers from four servers are sufficient to ensure the requested requirements.	
%
%
%

	
It is worth noting that our problem is performed for a snapshot assuming all the packets of the services which are generated in unit time are fetched into the network at the beginning of each unit time.
	%
		  Hence, the arrival time of all packets is the same and can be set to zero.
		   Therefore, 
	the total service chain delay for each user $u$ at BS $j$ on the requested service is inferred in \eqref{maxd}  \cite{game-theroy} in which $\hat B_{n,n'}$ is  bandwidth of link between node $n'$ to node $n$.
\begin{figure*}
	\begin{align}
	&D^{\text{Total}}_{u,j}=\max\limits_{\scriptstyle\forall n \in {\cal N},v \in {\cal V},\hfill\atop
		\scriptstyle f_m^s \in {\Omega _s},s \in {\cal S}\hfill}\Big\{\beta_{u,j,n}^{f_{m}^{s},v}e_v^{f_m^s}\Big  ( {t_{u,j,n}^{f_{m}^{s},v}}+\tau_{u,j,n}^{f_{m}^{s},v}+\sum_{n' \in \mathcal{N}}\dfrac{\sum_{ v' \in \mathcal{V}}\beta_{u,j,n'}^{f_{m-1}^{s},v'}e_v^{f_m^s}y_{u,j}}{\hat B_{n,n'}}  \Big)\Big \}, 
	\forall u \in \mathcal{U}.\label{maxd}
	\end{align}
	\hrule
\end{figure*}
Based on the delay requirement of each service, we have the following constrain:
\begin{align}
\label{delay}
\text{C8:~}D^{\text{Total}}_{u} \le D^{\text{max}}_{s},~\forall u\in\mathcal{U}. \nonumber
\end{align}
\subsection{\textbf{Objective Function and Optimization Problem}}
Our aim is to minimize the total energy per service in the network. To this end, we divide the total energy into three categories: 1) Radio part energy consumption, 2) CPU energy consumption, and 3) memory energy consumption. Therefore, the energy of the radio part is calculated as follows:
where $\tilde \tau$ is time unit, in which the optimization problem is solved. Moreover, the energy consumption of the CPU of servers consists of two parts: the energy consumption of the connected mode (when the server is working and host one or several VMs), the energy consumption of the idle mode (when the server is in sleep mode and no VMs run on it). Therefore, we have: 
\begin{align}
&E(\bold{B}, \boldsymbol{\tau})^{\text{CPU}}=\sum\limits_{\scriptstyle\forall n \in {\cal N},v \in {\cal V}, u \in \mathcal{U}\hfill\atop
	\scriptstyle f_m^s \in {\Omega _s},s \in {\cal S}, j \in \mathcal{J}\hfill}P_{v,n}^{\text{CPU}}\beta_{u,j,n}^{f_{m}^{s},v}e_v^{f_m^s}\tau_{u,j,n}^{f_{m}^{s},v}+\nonumber\\&
 \sum_{n \in \mathcal{N}} \hat P_n^{\text{CPU}}(\tilde \tau-\sum\limits_{\scriptstyle\forall v \in {\cal V}, u \in \mathcal{U}\hfill\atop
	\scriptstyle f_m^s \in {\Omega _s},s \in {\cal S}, j \in \mathcal{J}\hfill}\beta_{u,j,n}^{f_{m}^{s},v}e_v^{f_m^s}\tau_{u,j,n}^{f_{m}^{s},v}),
\end{align}
{Therefore, the total energy consumption function is given as follows \cite{7279063}:  
	\begin{align}\label{16}
	&\Psi(\bold{P}, \boldsymbol{\rho}, \bold{B}, \boldsymbol{\tau})=\mu_1 E(\bold{P}, \boldsymbol{\rho})^{\text{Radio}}+\mu_2E(\bold{B}, \boldsymbol{\tau})^{\text{CPU}}
	\end{align}
	where  $\mu_1$ and $\mu_2$ are constants and are used for scaling and balancing the energy of different resource types. }

\textcolor{black}{The gain of this approach is not only saving the consumption of the power in the active mode (i.e., under load) but also saving the power of the servers in the idle mode.} 
\subsection{\textbf{Problem Formulation}}  \label{Problemformulation}
Based on these definitions, our aim is to solve the following JRN-RA optimization problem: 
\begin{subequations}\label{MDPG}
	\begin{align}\label{op11}
		&
\max_{\bold {P}, \boldsymbol{\rho},\boldsymbol{\mathcal{\beta}}}\frac{\sum_{u\in\mathcal{U}}\sum_{j\in\mathcal{J}}\sum_{k\in\mathcal{K}} r_{u,j}^{k}}{\Psi(\bold{P},\boldsymbol{\rho}, \bold{B}, \boldsymbol{\tau})},
	\\
		\textbf{ s.t:~~}&\text{C1-C8 } \\
	&\label{1rhod}
	\rho_{u,j}^{k}\in \{0,1\},~\forall u\in\mathcal{U},j \in \mathcal{J},~k\in\mathcal{K},\\&
	\label{1betad}
	\beta_{u,j,n}^{f_{m}^{s},v} \in \{0,1\},\forall u\in\mathcal{U},\,\forall f_{m}^{s}\in\Omega_{s},\forall s\in\mathcal{S},
	\\& 
	\label{1xd}
	x_{u,j,n,u',j',n'}^{f_{m}^{s},v,f_{m'}^{s'},v'}\in \{0,1\},\forall u,u'\in\mathcal{U},u\neq u', \\&\forall f_{m}^{s}, f_{m'}^{s'}\in\Omega_{s'}, \forall v , v' \in \mathcal{V}, n,n' \in \mathcal{N},\nonumber
	\end{align}
\end{subequations}
where $\boldsymbol{\rho}=[\rho_{u,j}^{k}]$, $\boldsymbol{\mathcal{\beta}}=[\beta_{u,j,n}^{f_{m}^{s},v}]$, 
 $\bold{{P}}=[p_{u,j}^{k}] $. 
In problem \eqref{op11},  constraints C1-C8 are defined in the previous section. 
Constraints \eqref{1rhod}-\eqref{1xd} are  for binary variables.

\section{MARKOV DECISION PROCESS BASED PROBLEM FORMULATION}\label{solutions}
	Due to the arrival of unpredictable users' requests, the state of the network, traffic, and capacity of the servers will be changed and will not be static. Thus, we are looking for a dynamic and appropriate solution to solve our proposed system.
	In this paper, we want to perform different NFs on the different VMs and assign VMs to the server/physical nodes. Since these choices are not specific and may have high dimensional state spaces, we adopt deep RL approaches for the joint radio and core resource allocation in the NFV-based network from the EE perspective.
	Hence, we investigate deep learning methods. Since we have compounded actions, such as continuous and discrete variables, we propose a soft actor-critic (SAC) algorithm modeled as an MDP. The elements of the MDP, i.e., state, action, and reward, are defined as follows:

	For each time step $t_p$, the state is forwarded for the actor and critic. The actual policy takes the state $s_{t_p}$ and outputs an action $a_{t_p}$, which results in a new stat $\tilde{s}_{t_p}$ and a reward $r_{t_p}$. 
	\subsection{System States:}
The system state is an abstraction of the environment, and the learner makes action decisions based on the states. The most influential parameter on the state of the network environment is channel gain.  
Therefore, the system state $\mathcal{S}$ is defined as is the state space, and
the state $s_{t_p} \in \mathcal{S}$ at $t_p$-th time slot that is equivalent to channel gain of all users at $t_p$-th time slot is represented as:
	\begin{align}\label{state}
	s_{t_p}=\{h_{u,j}^{k}(t_p)\}.
	\end{align}
	\subsection{Action space:}
	The learner takes an action by considering the network's state because the network operates in a new state and transients from the current state of the network. The set of action space $\mathcal{A}$, which $a_{t_p} \in \mathcal{A}$ executed in the ${t_p}$-th time slot is expressed as:
		\begin{align}
	a_{t_p}=\{P(t_p),\rho(t_p),\beta(t_p)\},
		\end{align}
		where, $P(t_p)$, $\rho(t_p)$, and $\beta(t_p)$ are the corresponding actions of state that mentioned before in time slot $t_p$. 
\subsection{Reward Function:}
	After an action is taken, the environment will return an immediate reward $r_{t_p}$ to the agent. In this system, the agent tries to improve the optimization problem \eqref{MDPG}, with the aim of maximization of EE. The benefit of the action is defined as the reward $r_{t_p}$. Thus, the reward is denoted as:
	\begin{align}
	\label{reward}
	r_{t_p}=c\frac{\sum_{u\in\mathcal{U}}\sum_{j\in\mathcal{J}}\sum_{k\in\mathcal{K}} r_{u,j}^{k}}{\Psi(\bold{P},\boldsymbol{\rho}, \bold{B}, \boldsymbol{\tau})}, 
	\end{align}
	where $c$ is a coefficient factor. 
Based on the actions that users take according to the state of the system, the system's state will be changed. For instance, if the learning solution increases the objective function, the reward will be positive; otherwise, the reward will be negative.
\textcolor{black}{\section{PROBLEM SOLUTION WITH SOFT ACTOR-CRITIC DRL ALGORITHM}
To solve our proposed MDP problem, we employs SAC DRL. The SAC algorithm consists of the following four parts:
\begin{itemize}
	\item Actor: Separate policy DNN,
	\item Critic: value function DNNs,
	\item An off-policy method which enables reuse of the past experience stored in the memory $\mathcal{M}$,
	\item Entropy maximization to ensure stability.
\end{itemize}
}
\subsection{Soft Value Functions}
The main objective of the RL agent is to find the policy $\pi(a_{t_p}\mid s_{t_p})$ to maximize the expected reward. $\pi$ is the parameter to show the stochastic policy that gives the probability of taking a specific action when the agent is in a certain state.
 Despite existing RL algorithms, to ensure that the agent can explore continually, an entropy term $-\log\pi(a_{t_p}\mid s_{t_p})$ is added to the reward in  \cite{17reinforcement}. 
 \textcolor{black}{The objective with the expected entropy of the policy $\pi(a_{t_p}\mid s_{t_p})$, which is the parameter to show the stochastic policy, gives the probability of taking a certain action when the agent is in a certain state is termed as the entropy objective and written as:}
\begin{align}\label{Valuefunc}
V^{\pi}=\mathbb{E}\{{\sum_{i=1}^{T^{\prime}}}\nu^{t_p} [r_{t_p} -\lambda \log\pi(a_{t_p}\mid s_{t_p})]\mid \pi\}
\end{align}
where $\nu^{t_p}$ shows the discount factor. Moreover, $\nu \in [0,1]$ and  $\lambda$ are the parameters that responsible for controlling the stochasticity of the optimal policy. 
 Furthermore, $\tilde{Q}^{\pi}$ is the estimate of expected reward if the action $a_{t_p}$ in state $s_{t_p}$ according to the policy $\pi$ was taken. Q-value function:
\begin{align}\label{tildeQ}
\tilde{Q}^{\pi}(s_{t_p},a_{t_p})=\mathbb{E}\{{\sum_{i=1}^{T^{\prime}}}\nu^{\pi} [r_{t_p} -\lambda \log\pi(a_{t_p}\mid s_{t_p})]\mid a_{t_p}, s_{t_p}, \pi\}
\end{align}
 The relevance of state-value function and Q-function is:
 	\begin{align}\label{anteg}
 	V^{\pi}(s_{t_p})=\lambda \log \int_{A} \exp (\frac{1}{\lambda}\tilde{Q}^{\pi}(s_{t_p},a_{t_p}))da
 	\end{align}
 	and the optimal policy for \eqref{Valuefunc}, according to the state and action that was chosen is defined as below:
\begin{align}\label{aast}
\pi^{\ast}(0 \mid s_{t_p})=\text{exp}( \frac{1}{\lambda}(\tilde{Q}^{\pi}(s_{t_p},a_{t_p}) - V^{\pi}(s_{t_p})))
\end{align}
Compared to the conventional actor-critic DRL algorithm \cite{conventional}, and that the \eqref{anteg} is a $LogSumExp$ function which is a soft maximum, the Q-value function
and the state-value function are also named as soft Q-value
function and soft state-value function, respectively.
\subsection{Critic Networks:}
In this section, among many effective methods \cite{sutton2018reinforcement}, we apply DNN to approximate value function due to its convergence and stability \cite{sutton2018reinforcement}.
According to the goal of policy evaluation, soft Q-value $Q_{w_i}(s_{t_p},a_{t_p})$ is approximated by a DNN which is symbolized by weight $w_i$. In this regard, we use memory  $\mathcal{M}$, in which the new generated experience at each time step is stored, to increase system stability and elude the divergence. In this process, $w_i$ is updated from the memory  $\mathcal{M}$ to fracture the correlations between samples.  Hence, the loss function can be expressed as follows:
\begin{align}
L(w_i)=\mathbb{E}[\frac{1}{2}(\tilde{Q}_{w_i}(s_{t_p},a_{t_p}\mid w_i)-H_{w_i}(s_{t_p},a_{t_p}))^2],
\end{align}
where term $H_{w_i}(s_{t_p},a_{t_p})$ is:
\begin{align}
H_{w_i}(s_{t_p},a_{t_p})=r_{t_p}+\nu V^{\pi}.
\end{align}
As previously explained, to update the weight of critic network, i.e, $w_i$, with gradient decent method to minimize the loss function $L(w_i)$, we have: 
\begin{align}
w_i(new)=w_i(old)-\alpha_{w_i}\nabla_{w_i} L(w_i(old)),
\end{align}
where $\alpha_{w_i}$ is the learning rate of the critic which is a positive value and $\nabla_{w_i} L(w_i)$ is the gradient of $L(w_i)$ and achieved by:
\begin{align}
\nabla_{w_i}L(w_i)=&\\
&\nabla_{w_i}\tilde{Q}_{w_i}(s_{t_p},a_{t_p})[\tilde{Q}_{w_i}(s_{t_p},a_{t_p})-H_{w_i}(s_{t_p},a_{t_p})]\nonumber.
\end{align}
\subsection{Actor Networks:}\label{Actor Networks}
In this part based on a policy improvement approach, the objective function can be written as:
\begin{align}
L(\delta_i)=\mathbb{E}[-\frac{1}{2}(\lambda \log \pi _{\delta_i}(a_{t_p} \mid s_{t_p})+H_{\delta_i}(s_{t_p},a_{t_p}))^2],
\end{align}
where term $H(s_{t_p},a_{t_p})$ is:
\begin{align}
H_{\delta_i}(s_{t_p},a_{t_p})= Q_{\pi}(s_{t_p},a_{t_p})-V^{\pi}(s_{t_p}).
\end{align}
\textcolor{black}{In a similar way of critic network, to update the weight of actor network $\delta_i$ with gradient decent method to minimize the loss function $L(\delta_i)$, we have:}  
\begin{align}
\delta_i(new)=\delta_i(old)+\alpha_{\delta_i}\nabla_{\delta_i} L(\delta_i(old))
\end{align}
where $\alpha_{\delta_i}$ is the learning rate of the actor network and $\nabla_{\delta_i} L(\delta_i)$ is the gradient of $L(\delta_i)$ and achieved by:
\begin{align}
\nabla_{\delta_i}L(\delta_i)=&\\
&\nabla_{\delta_i}V^{\pi}(s_{t_p})[-H(s_{t_p},a_{t_p})+\lambda\log \pi _{\delta_i}(a_{t_p})]\nonumber.
\end{align}
\subsection{The Soft Actor-Critic DRL Algorithm}
The duty of agent is divided as follows:
\begin{itemize}
 \item The actor network: mapping states to actions,
 \item The critic network: estimating states and state-action couple,
 \item The memory $\mathcal{M}$: storing experience.	
\end{itemize}
The actor network with the suitable approach of finding the right policy $\pi_{\delta_i}(0 \mid s_{t_p})$, which will be trained based on \eqref{aast}, according to the current state$s_{t_p}$ selects and executes an action $a_{t_p}$.  
\textcolor{black}{In this step the neural network, depending on whether the action is well chosen or not, the agent receive rewards or punishments, respectively.} Then, the next state ${\tilde{s}}_{t_p}$ is characterized and memory  $\mathcal{M}$ store this tuple $(s_{t_p},a_{t_p},r_{t_p},{\tilde{s}}_{t_p})$.
\textcolor{black}{The critic network samples randomly from the tuples that are stored in the memory  $\mathcal{M}$, with the approach of elimination of temporal correlations between samples and train the state-value and Q-value.} This learning process is intermittent sampling of the environment and updating the network until the final state or the network time-out.
To understanding the better concept of this cycle, we propose Algorithm \ref{algorithmic}.
\begin{algorithm} 
	\small
	\caption{Soft Actor-critic Algorithm \cite{SoAcCrDRL}}\label{algorithmic}
	Initialize parameters $w_i, \delta_i$ and empty memory $\mathcal{M}$ 
	\\
	Set target parameters equal to main parameters
	\\
	$\mathbf{for}$ each iteration
	~~~~~~~~~~~~~~~~~~~~~~~~~~~~~~~~~~~~~~~~~~~~~~~~~~~~~~~~~~~~~~~
	$\mathbf{for}$ each environment step:
	~~~~~~~~~~~~~~~~~~~~~~~~~~~~~~~~~~~~~~~~~~~~~~~~~~~~~~~~~~~~~~~
	Observe state $s_{t_p}$ and select action $a_{t_p} \sim \pi(0 \mid s_{t_p})$
	~~~~~~~~~~~~~~~~~~~~~~~~~~~~~~~~~~~~~~~~~~~~~~~~~~~~~~~~~~~~~~~
	Execute $a_{t_p}$ in the environment
	~~~~~~~~~~~~~~~~~~~~~~~~~~~~~~~~~~~~~~~~~~~~~~~~~~~~~~~~~~~~~~~
	Observe next state ${\tilde{s}}_{t_p}$, reward $r_t$
	~~~~~~~~~~~~~~~~~~~~~~~~~~~~~~~~~~~~~~~~~~~~~~~~~~~~~~~~~~~~~~~
	Store $(s_{t_p},a_{t_p},r_{t_p},{\tilde{s}}_{t_p})$ in memory and update $\mathcal{M}\leftarrow\mathcal{M}\cup\{(s_{t_p},a_{t_p},r_t,{\tilde{s}}_{t_p})\}$
	~~~~~~~~~~~~~~~~~~~~~~~~~~~~~~~~~~~~~~~~~~~~~~~~~~~~~~~~~~~~~~~
    $\mathbf{end}$ $\mathbf{for}$
    ~~~~~~~~~~~~~~~~~~~~~~~~~~~~~~~~~~~~~~~~~~~~~~~~~~~~~~~~~~~~~~~
    $\mathbf{for}$ each gradient step:
    ~~~~~~~~~~~~~~~~~~~~~~~~~~~~~~~~~~~~~~~~~~~~~~~~~~~~~~~~~~~~~~~
    $w_i(new)\leftarrow w_i(old)-\alpha_{w_i}\nabla_{w_i} L(w_i(old))$
    ~~~~~~~~~~~~~~~~~~~~~~~~~~~~~~~~~~~~~~~~~~~~~~~~~~~~~~~~~~~~~~~
    $\delta_i(new)\leftarrow \delta_i(old)+\alpha_{\delta_i}\nabla_{\delta_i} L(\delta_i(old))$
    ~~~~~~~~~~~~~~~~~~~~~~~~~~~~~~~~~~~~~~~~~~~~~~~~~~~~~~~~~~~~~~~
    $\mathbf{end}$ $\mathbf{for}$
    	\\
      $\mathbf{end}$ $\mathbf{for}$
      \\
output parameters:  $w_i, \delta_i$
\end{algorithm}
\section{COMPUTATIONAL COMPLEXITY AND CONVERGENCE ANALYSIS}\label{Complexity-Convergance}
\subsection{Computational Complexity}
The proposed system's computational complexity is one of the key and practical criteria of the proposed system. Thus,  we  analyze  the computational complexity of our proposed system model from two perspectives: training process and running process.
\begin{itemize}
	\item Training Process:
In most of the papers such as \cite{8932445},\cite{81702199999}, the authors ignored the computation complexity of the training process and assumed that the training process happens in off-line mode, and its computation complexity is proportional to the size of the training data and duration of the training. We consider the number of neurons on $l$th and $k$th layer of the actor and critic network as $I^l$ and $V^k$,  respectively. Hence, the total computational complexity of the training process is $\mathcal{O}(\sum_{l=2}^{L-1}(I^{l-1}I^l+I^lI^{l+1})+\sum_{k=2}^{K-1}(V^{k-1}V^k+V^kV^{k+1}))$\cite{817021999}.
	\item Running Process:
In this part, we propose the computational complexity of the actor critic network. 
 The internal computations of neural network algorithms depends on the architecture, the number of neurons, and the number of layers. Besides, in an algorithm consistent with deep RL, the number of states and actions is related to the number of neurons and neural network layers. Therefore, the computational complexity is in order of $\mathcal{O}(B\times(2 \times HZ)\times(J\times U\times S\times V\times M))$ in which $B$ is the training batch, $H$ is the number of neurons in hidden layer $Z$, $J$, $U$, $S$, $V$, and $M$ are the numbers of BSs, users, NS, VMs, and NFVs, respectively. 
\end{itemize}
\subsection{Convergence Analysis}

In this section, we  evaluate the convergence of the SAC algorithm. According to Q-function in \eqref{tildeQ}, the solution can converge to the optimal Q-function as $t \to \infty $ with probability 1, if the learning rates of the algorithm $\alpha_{w_i}$ and $\alpha_{\delta_i}$ are deterministic, non-increasing, and satisfy \cite{grondman2012survey}:
\begin{align}
\sum_0^{\infty} \alpha_{w_i}=\infty,~~~\sum_0^{\infty} (\alpha_{w_i})^2 \textless \infty, \\ \sum_0^{\infty} \alpha_{\delta_i}=\infty,~~~\sum_0^{\infty} (\alpha_{\delta_i})^2 \textless \infty,~~~\lim_{t_p\to \infty}\frac{\alpha_{\delta_i}}{\alpha_{w_i}}=0,
\end{align}
and $\mid r_{t_p} \mid$ be bounded\cite{817021888}. \textcolor{black}{With the aim of fast convergence and training our neural network effectively, we utilize the inverse time decaying learning rate that uses the large learning rate in the first episodes to prevent the network from getting stuck in a bad local optimum trap near the initial point and uses the small learning rate in the last training epochs in order to converge to a good local optimum \cite{8170218888}.} The convergence of our proposed algorithm is also analyzed through simulations results in Section \ref{simulations}.
\begin{algorithm} 
	\small
	\caption{Actor-critic methods consist of action-value $\tilde{Q}$ or state-value $V_{\Pi}$}\label{alg1}
	Initialize randomly $s , \theta, w$
	\\
	For $t=1,\cdots,T$: ~~~~~~~~~~~~~~~~~~~~~~~~~~~~~~~~~~~~~~~~~~~~~~~~~~~~~~~~~~~~~~~
	1.Sample reward $r_{t_p}~R_{t_p}(s,a)$ and next state $s^{\prime}~P(s^{\prime} \mid s,a)$;
	~~~~~~~~~~~~~~~~~~~~~~~~~~~~~~~~~~~~~~~~~~~~~~~~~~~~~~~~~~~~~~~
	2.Then sample the next action $a^{\prime}~ \Pi_{\theta}(a^{\prime} \mid s^{\prime})$;
	~~~~~~~~~~~~~~~~~~~~~~~~~~~~~~~~~~~~~~~~~~~~~~~~~~~~~~~~~~~~~~~
	3.Update the policy parameters: $\theta_i (new)\leftarrow \theta_i (old)+a_{\theta}\nabla J_i(\theta_i(old)) $;
	~~~~~~~~~~~~~~~~~~~~~~~~~~~~~~~~~~~~~~~~~~~~~~~~~~~~~~~~~~~~~~~
	4.Compute the correction (TD error) for action-value at time $t_p$:
	$\delta_{t_p}=\alpha[R_{s\rightarrow s^{\prime}} + \nu \max \tilde{Q}_{t_p}(s^{\prime},a^{\prime})-\tilde{Q}(s,a)]$
	~~~~~~~~~~~~~~~~~~~~~~~~~~~~~~~~~~~~~~~~~~~~~~~~~~~~~~~~~~~~~~~
	5.and use it to update the parameters of action-value function:
	$w_i(new)=w_i(old)+a_w\nabla L(w_i(old))$\\
	Update $a \leftarrow a^{\prime}$ and $s \leftarrow s^{\prime}$.
\end{algorithm}
\begin{algorithm} 
	\small
	\caption{Join Radio and Core Resource Allocation}
	\label{alg2}
	Initialize total reward $r_t=0$\\
	\For{ $t=1,\cdots,T$}{\For{$u=1,\cdots,U$ }{The agent take an actoion $a$ with considering state $s$.
			\\ 
			\If{ constarints C1-C8 }{Calcute the obtained rate for user $u$\\
				Calcute the EE by \ref{reward}}
		}
	}
\end{algorithm}
\begin{algorithm} 
	\small
	\caption{Actor-critic methods consist of action-value $\tilde{Q}$ or state-value $V_{\Pi}$}\label{alg1}
	Initialize randomly $s , \theta, w$
	\\
	For $t=1,\cdots,T$: ~~~~~~~~~~~~~~~~~~~~~~~~~~~~~~~~~~~~~~~~~~~~~~~~~~~~~~~~~~~~~~~
	1.Sample reward $r_{t_p}~R_{t_p}(s,a)$ and next state $s^{\prime}~P(s^{\prime} \mid s,a)$;
	~~~~~~~~~~~~~~~~~~~~~~~~~~~~~~~~~~~~~~~~~~~~~~~~~~~~~~~~~~~~~~~
	2.Then sample the next action $a^{\prime}~ \Pi_{\theta}(a^{\prime} \mid s^{\prime})$;
	~~~~~~~~~~~~~~~~~~~~~~~~~~~~~~~~~~~~~~~~~~~~~~~~~~~~~~~~~~~~~~~
	3.Update the policy parameters: $\theta_i (new)\leftarrow \theta_i (old)+a_{\theta}\nabla J_i(\theta_i(old)) $;
	~~~~~~~~~~~~~~~~~~~~~~~~~~~~~~~~~~~~~~~~~~~~~~~~~~~~~~~~~~~~~~~
	4.Compute the correction (TD error) for action-value at time $t_p$:
	$\delta_{t_p}=\alpha[R_{s\rightarrow s^{\prime}} + \nu \max \tilde{Q}_{t_p}(s^{\prime},a^{\prime})-\tilde{Q}(s,a)]$
	~~~~~~~~~~~~~~~~~~~~~~~~~~~~~~~~~~~~~~~~~~~~~~~~~~~~~~~~~~~~~~~
	5.and use it to update the parameters of action-value function:
	$w_i(new)=w_i(old)+a_w\nabla L(w_i(old))$\\
	Update $a \leftarrow a^{\prime}$ and $s \leftarrow s^{\prime}$.
\end{algorithm}
\begin{algorithm} 
	\small
	\caption{DDPG Algorithm}
	\label{alg_DDPG}
	\textbf{Initialization:} Set the soft update parameter as $\tau=0.05$, the discount factor as $\gamma=0.9$, the maximum memory size as $|\mathcal{M}|=100000$, epsilon for exploration as $eps=1$, and decay epsilon factor $eps_{decay}=0.9994$.\\
	\For{ $t=1,\dots,E$}{Select action $a_{t}=\pi(S_t|\theta^{\pi})+n_t$\\Take the action $a_t$ and observe the reward $r_t$ and observe the new state $s_{t+1}$
	\\Store transaction $<s_t,a_t,r_t,s_{t+1}>$ in momory $\mathcal{M}$\\
Let $y_t=r_t+\gamma$\\
Update the critic network and minimize the loss function:\\
$\nabla_{\theta}J=\frac{1}{N}\sum_{i}\nabla_{a}Q(s,a|\theta^{Q})|_{s=s_t,a=\pi(s_t)}\nabla_{\theta}\pi(s|\theta^{\pi})|_{s_t}$\\
Update the target network parameters:
\\
$\theta^{Q'}\leftarrow\rho\theta^{Q}+(1-\rho)\theta^{Q'}$\\
$\theta^{\pi'}\leftarrow\rho^{\theta^\pi}+(1-\rho)\theta^{\pi'}$}
\end{algorithm}

\section{SIMULATION RESULTS}\label{simulations}
In this part, we consider a square area $1000 m \times 1000 m$ with 4 BSs, and the users are randomly distributed there. Moreover, we assume 10 subchannels with the frequency bandwidth of 200 kHz, the maximum power of each BS is set to 40 watts, and the minimum data rate of the downlink is set to 1 bps/Hz.    In addition, our proposed system has 5 VNFs. 
Since each VNF require different processing parameters as $\pi^{user}$, we define some VNFs in Table \ref{PROCESSING REQUIREMENT (PER USER) FOR THE VNFS} \cite{861130000}. Moreover, we represent three types of SFCs by changing the order of six VNFs in Table \ref{PERFORMANCE REQUIREMENTS FOR THE SFCS}. Also, we show the performance requirements in terms of bandwidth $\delta$ and maximum tolerated delay $\varphi$ for all SFCs. The upper bound for the total service chain delay is set between 75 to 100 ms \cite{8932445}. Furthermore, we assume that each node can host 6 VMs, and each VMs node can host 6 VNF at most. The capacities of bandwidth and memory of each physical link and node are set to $1$ Gbps and $1$ Gbyte, respectively. Moreover, the CPU capacity of each node is set to 1200 CPU cycle/bit \cite{ebrahimi2020joint}.  
\textcolor{black}{In the training process, the value of loss function is Mean Square Error (MSE).}  
From the commuting services point of view, we should explain that at the start of each experiment, we assume that each SFC request consists of six VNFs.  
\begin{table}[t]	
	\renewcommand{\arraystretch}{1.5}
	\centering
	\caption{Processing Requirement (per User) for the VNFs}
	\begin{adjustbox}{width=.48\textwidth,center}	
		\begin{tabular}{|c |c | }	
			\hline
			\textbf{Virtual Network Function}&$\pi^{user}$\\\hline
			Network Address Translator (NAT)&0.00092\\ \hline
			Firewall (FW)&0.0009\\ \hline
		Traffic Monitor (TM)&0.0133\\ \hline
			WAN Optimization Controller (WOC) &0.0054\\ \hline
		Intrusion Detection Prevention System (IDPS) &0.0107\\ \hline
		Video Optimization Controller (VOC)&0.0054\\\hline
		\end{tabular}
	\end{adjustbox}\label{PROCESSING REQUIREMENT (PER USER) FOR THE VNFS}
\end{table}  
\begin{table}[t]	
	\renewcommand{\arraystretch}{1.5}
	\centering
	\caption{Performance Requirement for the SFCs}
	\begin{adjustbox}{width=.48\textwidth,center}	
		\begin{tabular}{|c |c |c |c | }	
			\hline
			\textbf{SFC}&\textbf{Chained VNFs}&\textbf{Latency}$(\varphi)$&\textbf{Bandwidth}$(\delta^{user})$\\\hline
			Web Service&NAT-FW-TM-WOC-IDPS&500 ms&100 kb/s\\ \hline
			VoIP&NAT-FW-TM-FW-NAT&100 ms& 64 kb/s\\ \hline
			Video Streaming&NAT-FW-TM-VOC-IDPS&100 ms&4 Mb/s\\ \hline
		\end{tabular}
	\end{adjustbox}\label{PERFORMANCE REQUIREMENTS FOR THE SFCS}
\end{table}

It is worth noting that the source code of the proposed method is written with Python language with \textit{TensorFlow}  library \cite{xiao2019nfvdeep} with Adam optimizer. Also, more details for configuration of the neural network are summarized in Table \ref{setting_neural}. 
Besides, the sources codes of the proposed method and baselines are available in \cite{bibid}.

\begin{table}[]
	\centering
	\caption{Simulation Parameters}		\label{setting_neural}
	\begin{tabular}{|c|c|}
		\hline
	\textbf{Parameter}	& 	\textbf{Value} \\ \hline
		Number of sub-carriers& 10  \\ \hline
	Total available bandwidth	& 200 KHz \\ \hline
    Number of  BSs	& 4 \\\hline
     Capacity of  transaction memory  & 100000\\\hline
   Batch size  & 32 \\\hline
   Actor network  learning rate&0.00001 \\\hline
      Critic network  learning rate&0.0001 \\\hline
         Moving average hyper parameter&0.05 \\\hline
         
        AWGN power & -170 dbm \\\hline
        Number of  hidden layer of DNN&2, 3, 4, 5, 6 \cite{xie2016service} \\\hline
        Number for neuron in each layer& 1024, 500, 250 \cite{xiao2019nfvdeep}\\\hline
	\end{tabular}
\end{table}

\subsection{Comparison between Joint Approach  and Disjoint	Approach }
\begin{figure*}[ht!]
	\centering
	\includegraphics[width=0.7\linewidth]{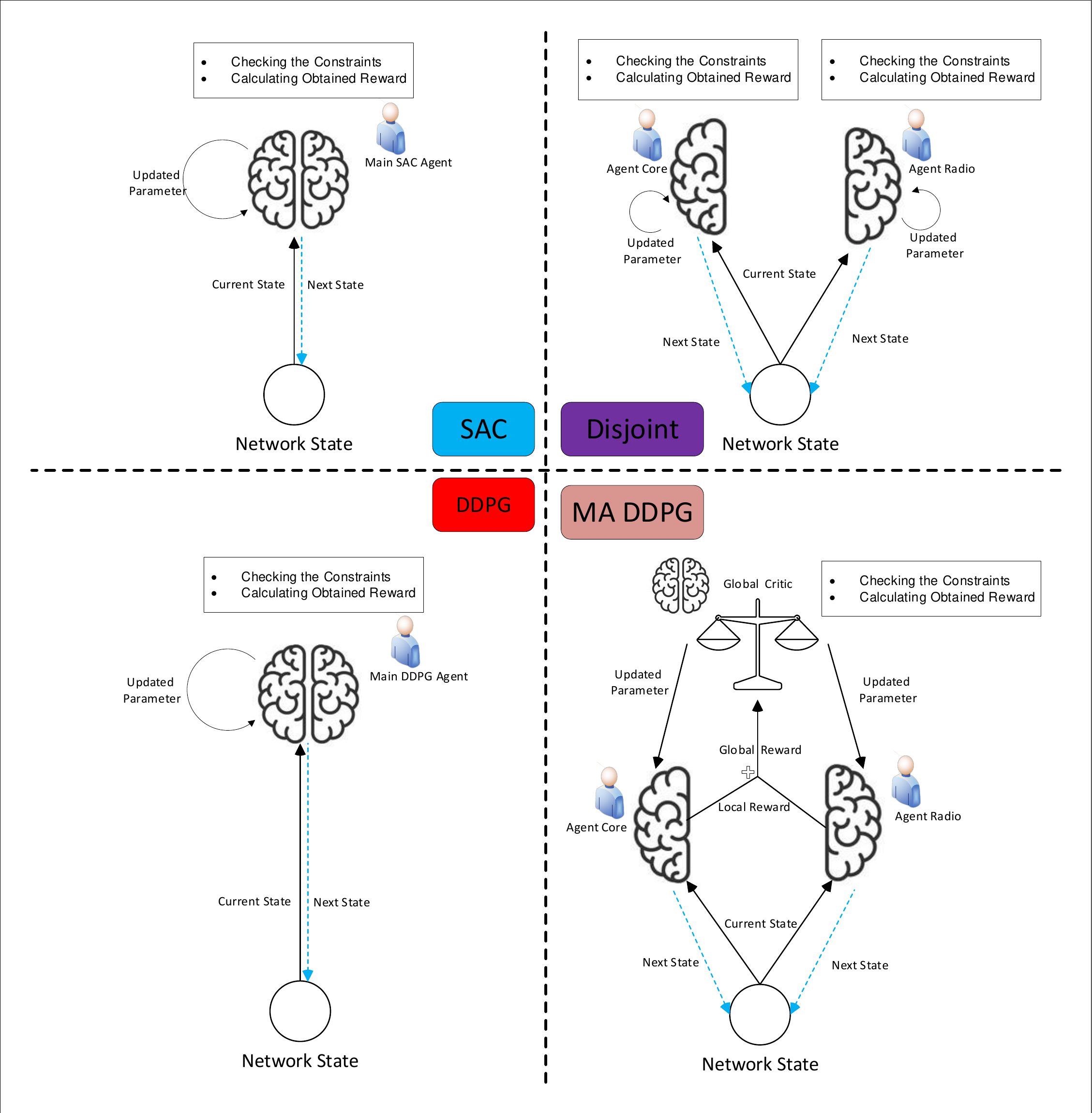}
	\caption{Comparing between the methods that are evaluated in this paper. In disjoint method, the core agent and radio agent selected their action based in the network state, independently. Similarly, in MA DDPG, each of the agents take the actions independently, but, they just report the obtained reward to the global critic. In addition the details of  SAC and DDPG methods are written in Al .\ref{alg1} and \ref{alg_DDPG}, respectively }
	\label{fig:comparision-method}
\end{figure*}

In this system model, we consider a joint radio and core resource allocation framework. In this part, we investigate the comparison between our proposed joint framework and disjoint framework in which the resource allocation optimization problem for RAN and core part are performed separately. To this end, we rewrite the problem formulation \eqref{16} as two separate problems in the following. Hence, the R-RA problem formulation can be written as:\                     
\begin{align}
	&\max_{\bold {P}, \boldsymbol{\rho},\boldsymbol{\mathcal{\beta}}}\frac{\sum_{u\in\mathcal{U}}\sum_{j\in\mathcal{J}}\sum_{k\in\mathcal{K}} r_{u,j}^{k}}{	\mu_1 E(\bold{P}, \boldsymbol{\rho})^{\text{Radio}}},\\ \nonumber
	&\textbf{s.t:~~} \text{C1-C2}.
\end{align}
This new optimization problem can be solved by SAC algorithm. On the other hand, NFV-RA optimization problem \eqref{16} can be summarized as follow: 
\begin{align}
	&\max_{\bold {P}, \boldsymbol{\rho},\boldsymbol{\mathcal{\beta}}}\dfrac{\sum_{u\in\mathcal{U}}\sum_{j\in\mathcal{J}}\sum_{k\in\mathcal{K}} r_{u,j}^{k}}{\mu_2E(\bold{B}, \boldsymbol{\tau})^{\text{CPU}}},\\ \nonumber
	&\textbf{ s.t:~~} \text{C3-C9}.
\end{align} 
We also solved the above  problem by SAC algorithm.  To solve above problems independently, first we solve R-RA problem. After that,  we consider the obtained rate  from radio part. Then we set the radio part delay as $10$ms. 
For the sake of evaluating the performance of the proposed SAC method, we  consider DDPG and multi-agent DDPG (MA DDPG) as baselines.\textcolor{black}{Table \ref{Complexity} shows the superiority of the SAC method, in terms of complexity, and the comparison with other method that considered as a baselines.}
\begin{table}[t]	
	\renewcommand{\arraystretch}{1.5}
	\centering
	\caption{Performance Complexity of SAC Compare to Baselines}
	\begin{adjustbox}{width=.24\textwidth,center}	
		\begin{tabular}{|c |c | }	
			\hline
			\textbf{Methods}&\textbf{Complexity}\\\hline
			SAC&$\mathcal{O}(B(2 \times HZ))$\\ \hline
			DDPG&..\\ \hline
		MADDPG&..\\ \hline
		\end{tabular}
	\end{adjustbox}\label{Complexity}
\end{table}
As we depict in Fig.\ref{fig:comparision-method}, it is assumed that the core network and radio part have independent agents in which the agents report the obtained reward to a central neural network called global critic network. 
 The global critic network calculates the total reward and critics between the agents. By updating the parameters of the radio and the core agents, these agent try to give better reward and minimize their loss function, gradually. It's worth noting that the global critic just have the global reward that is obtained from radio and core agents.

\begin{figure}
	\centering
	\includegraphics[width=1\linewidth]{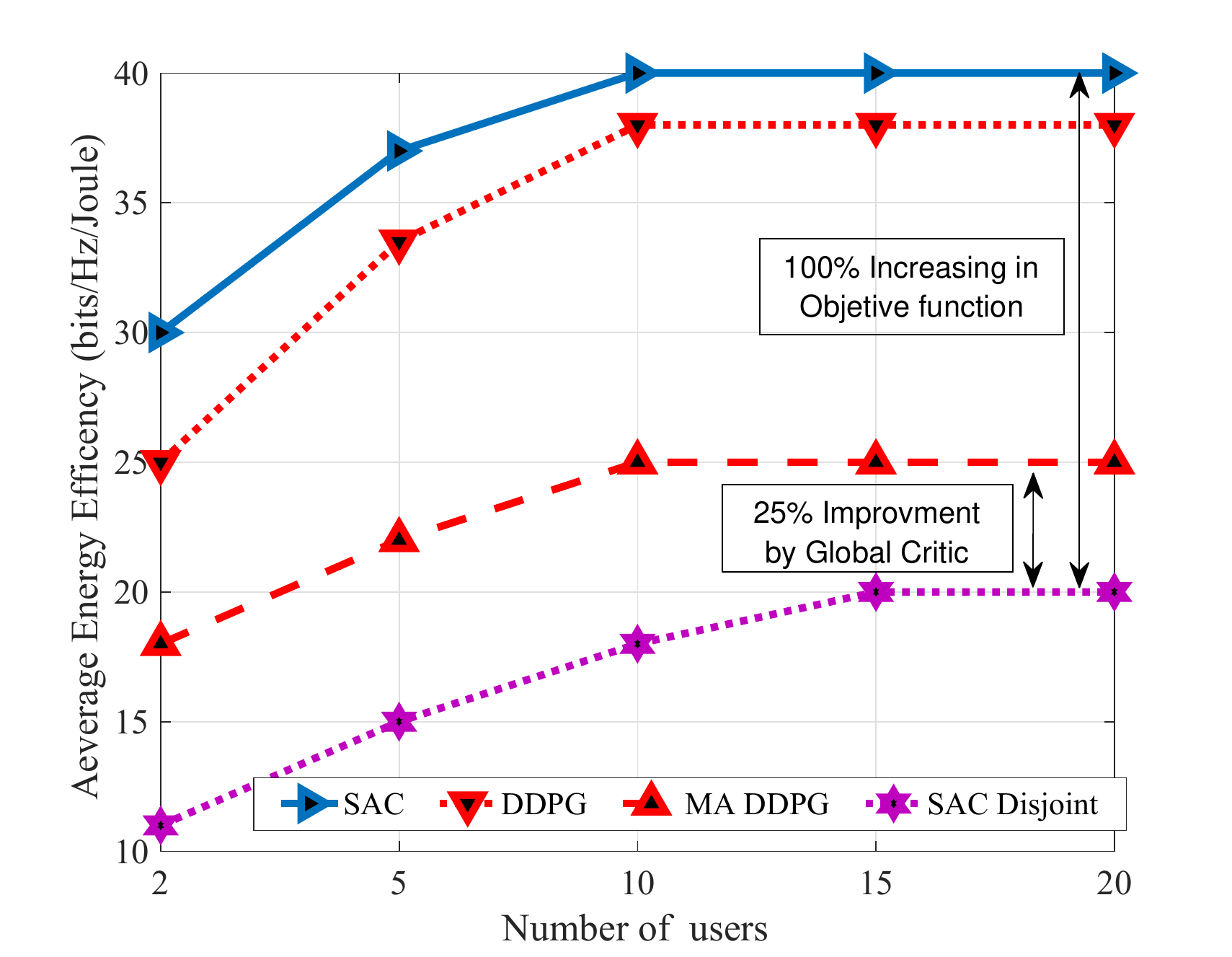}
	\caption{EE  for number of  the requested services}
	\label{fig:users}
\end{figure}

 As shown in Fig. \ref{fig:users}, EE in the disjoint approach is one-half of the joint approaches, i.e., SAC and DDPG. The reason for this is that in the disjoint approach, two problems are solved entirely independently of each other, and thus, we are faced with two problems that optimize the desired parts, e.g., RAN and core, regardless of the E2E view. As stated above, the global critic network calculates the total reward and critics between the agents. Therefore, it is expected that MA DDPG method has better performance than the disjoint approach which is revealed in Fig. \ref{fig:users}.

\subsection{Effects of Number of the Service Requests}
As can be seen in Fig. \ref{fig:users}, by increasing the number of users to 10 users, the EE increases with increasing number of users.  This is due to the fact that by  increasing number of users, the network consumes more energy to ensure new users' rate and delay requirements. Thus, the total data rate in the network is increased while the total energy is constant. 
Furthermore, as can be seen by increasing the number of users up to 20, the EE  remains constant. 
It stems from the fact that by increasing the number of users, we reach a point that all the network energy is consumed to guarantee the requirements of all users. Thus,  with the request of a new user, the network has no resources to guarantee the requirements of that user. As a result, the new user or one user who has a negative impact on EE is rejected. Therefore, the total data rate remains constant from one point in the figure onwards. The total network energy consumption is also constant and at its maximum value; hence, the EE remains constant.
\subsection{Effects of Minimum Data Rate }
\begin{figure}
	\centering
	\includegraphics[width=8.2 cm , height=6 cm]{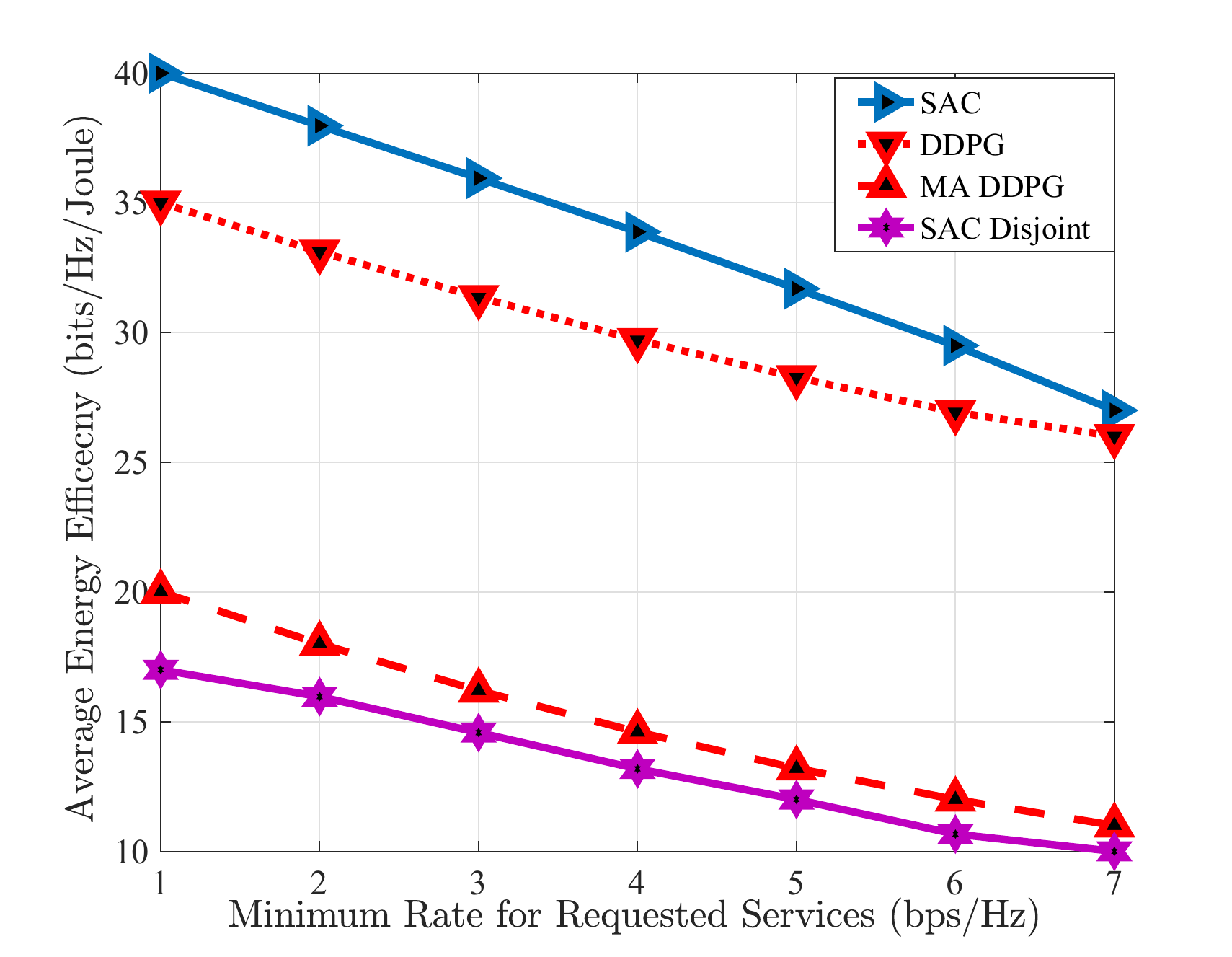}
	\caption{Average EE versus the minimum data rate of the requested services. }
	\label{fig:minrat}
\end{figure}

 As we show in Fig.\ref{fig:minrat}, the average EE is decreased by increasing the minimum data rate of the requested services. This is due to the fact that more energy is required to provide higher data rate. On the other hand, because of the resource  limitations, i.e., energy limitation, the total accepted user is decrees by increasing the minimum data rate of the services. Therefore, the EE is decreased by increasing the minimum rate requirement.

\subsection{Effects of Maximum Tolerable Time of the Services }
\begin{figure}
	\centering
	\includegraphics[width=8.2 cm , height=6 cm]{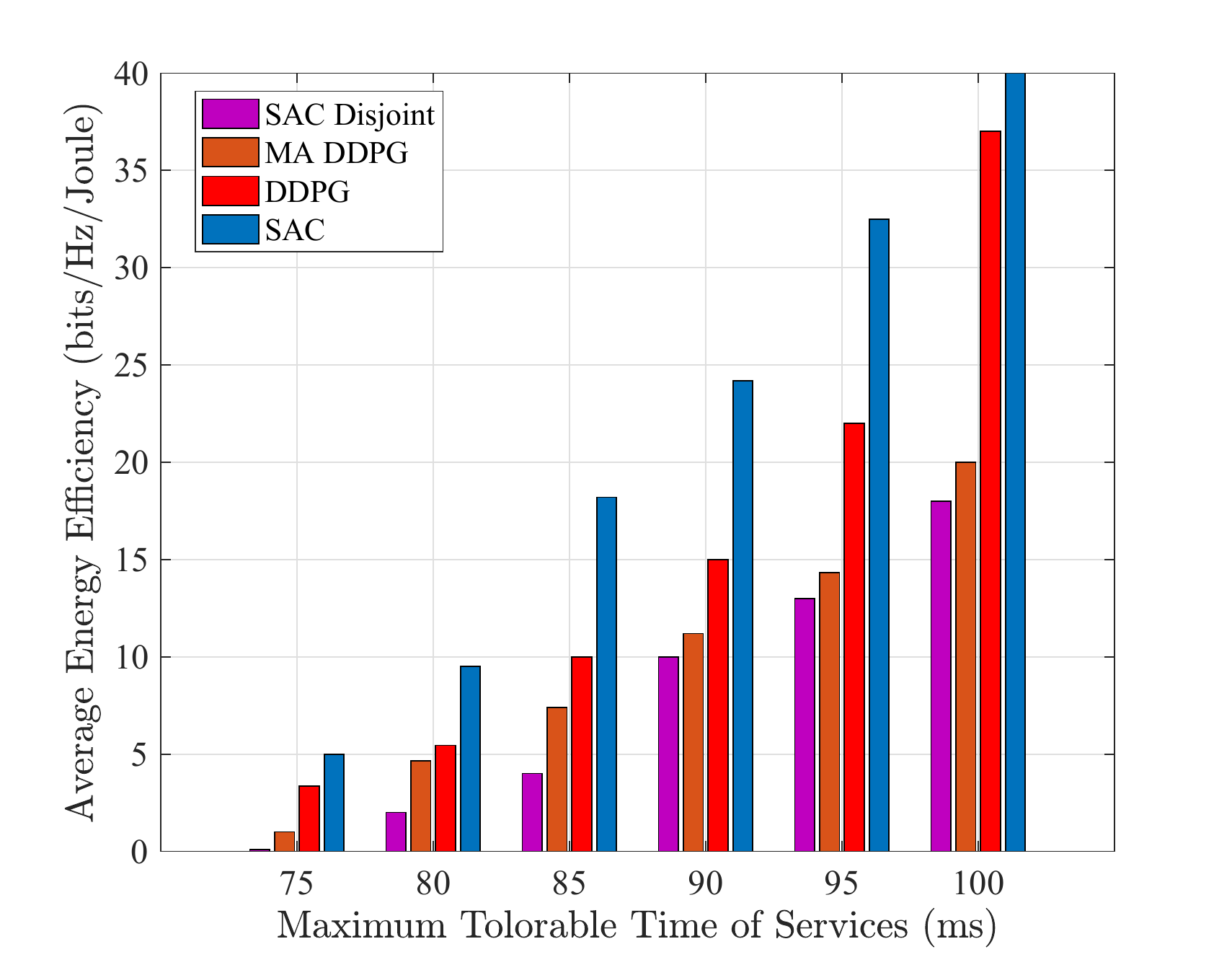}
	\caption{Average EE versus the maximum tolerable time of the requested services }
	\label{fig:maxdelay}
\end{figure}

Based on constraint (C8) and Eq. \eqref{maxd}, it is obvious that more energy is required to guarantee slighter delays, and as a result, fewer users can be accepted due to the limitation of energy resources. 
  As we depicted in Fig.\ref{fig:maxdelay}, by increasing the maximum tolerable time of the services, average EE of the network increases. Moreover, as stated before the disjoint approach performs worse than the other approaches, e.g., MA DDPG, DDPG, and joint SAC.

\subsection{Effects of the Coefficients}
\begin{figure}[ht!]
	\subfigure[Average EE of core versus coefficient core]{
	\includegraphics[width=8.2 cm , height=6 cm]{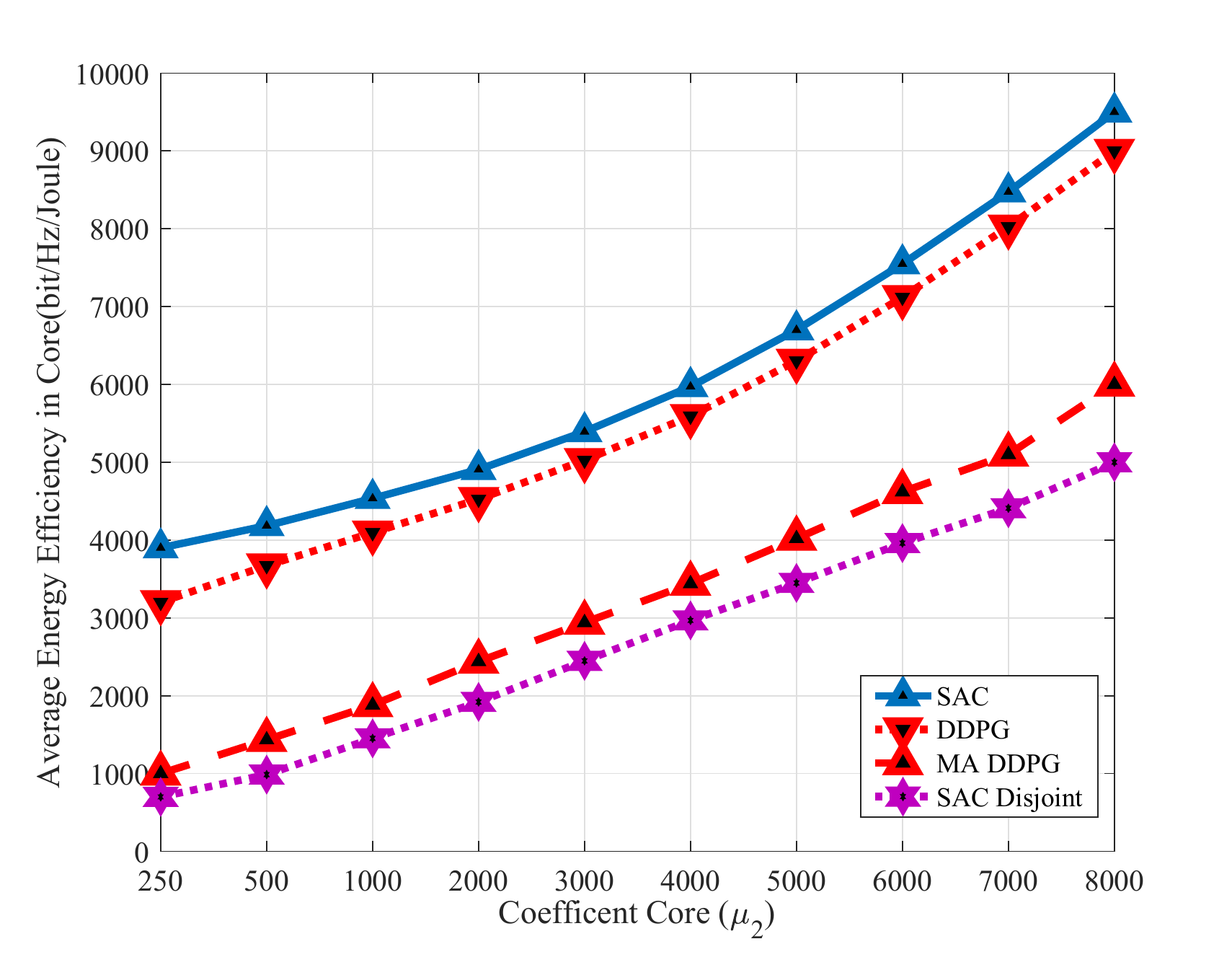}
	\label{Figure1}
	}
	\subfigure[Average EE of radio versus coefficient core]{
	\includegraphics[width=8.2 cm , height=6 cm]{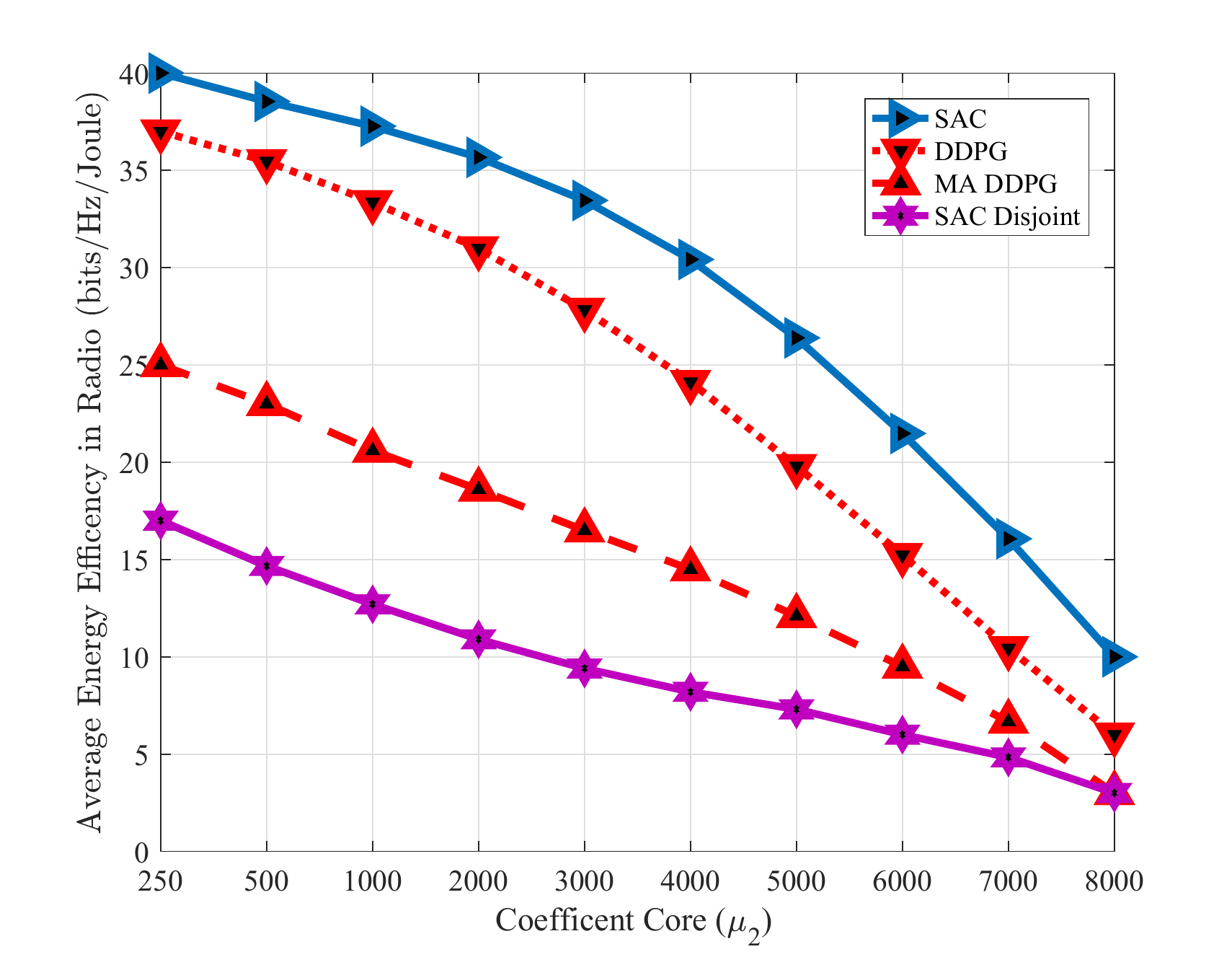}
	\label{Figure2}
	}
\subfigure[Average EE  versus coefficient core]{
\includegraphics[width=8.2 cm , height=6 cm]{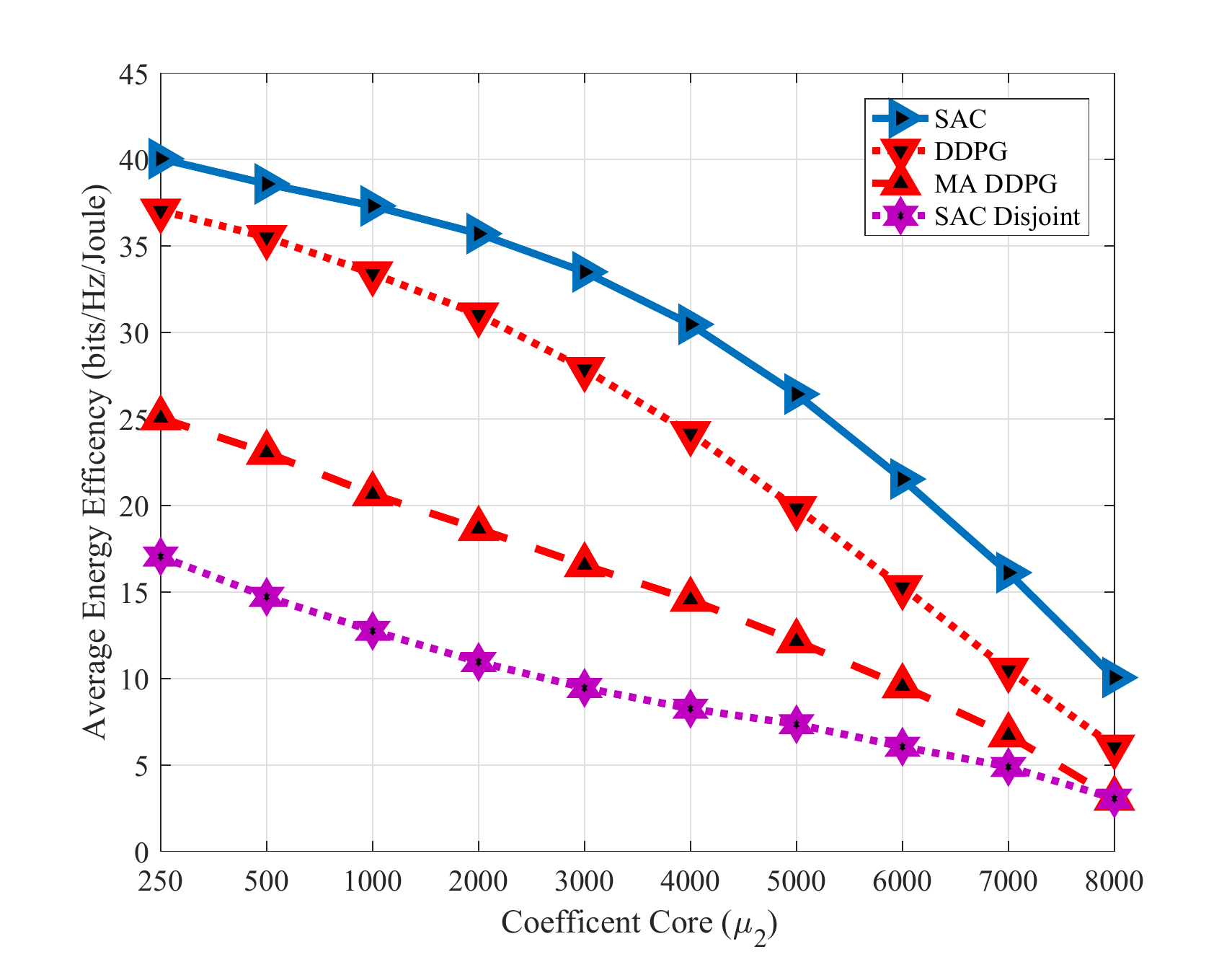}
\label{Figure3}}
	\caption{Average EE in core, radio, and both of them versus coefficient core.}\label{coefficent}
\end{figure}

As can be seen in Fig. \ref{coefficent}, by incensing the coefficient of core in the different methods, the agents try to maximize the EE in the core network.  As a result,  the agents select the VMs that have lower energy consumption.
 By selecting these VMs, there is the possibility that the physical path between the source node and destination can not guarantee the maximum tolerable delay of the services.
 On the other hand,  some users can not be admitted due to the limited computation resources in these VMs. 
Moreover, the total energy consumed in the core network is significantly less than the radio part; thus, the average EE in the network is very similar to the average efficiency in the radio part of the network, as is shown in Fig.\ref{Figure3} and Fig.\ref{Figure2}.
Furthermore, due to the lower energy consumption in the core network, the average EE in the core network is remarkably higher than EE in the radio part, as can be seen in Fig.\ref{Figure1}.
\subsection{Comparing  Signaling Overhead in the proposed method and the Baselines}

In each method, the agents need to have information such as the states and obtained rewards to take the actions.
Thus, this information must be intercommunicated between the resource orchestrator and the radio part and core network \cite{ng2011resource}. 
To measure and model this information,  we assume that each element of the channel gain matrix can be decoded as a fixed-length 16-bit binary string. Consequently, we use the type  '\textit{float16}' in \textit{Numpy} library in Python.  
Based on this, by considering a system  with $B$ BSs and $J$ subchannels and $U$ users, the size of the channel gain matrix in a binary model is equal to $16\times B\times J \times U$. Similarly, the local reward that each agent obtains can be modeled as a fixed-length 16-bit binary string. Finally, we summarize the total signal overhead in each episode for the proposed method and the baselines in Table \ref{overhead}. 
\begin{table*}[ht]
	\centering
	\caption{Total Signal overhead for the methods}
	\scalebox{.7}{
	\label{overhead}
	\begin{tabular}{|c|c|c|c|}
		\hline
		\textbf{Method} & \begin{tabular}[c]{@{}c@{}}\textbf{Size of the data that intercommunicates}\\ \textbf{between the radio part and core network}\end{tabular} & \begin{tabular}[c]{@{}c@{}}\textbf{Size of the data that intercommunicates} \\ \textbf{between the resource orchestrator and core network}\end{tabular} & \begin{tabular}[c]{@{}c@{}}\textbf{Size of the data that intercommunicates}\\ \textbf {between the resource orchestrator and radio part}\end{tabular} \\ \hline
	SAC Disjoint	&      Requested data rate in radio part=16 bits                &  Ignore able                                         &    Ignore able                           \\ \hline
	SAC, DDPG	&    Requested data rate in radio part=16 bits         &  16$\times B \times J \times U$    bits                   & 16$\times B \times J \times U$    bits                 \\ \hline
MA DDPG	&     Requested data rate in radio part=16 bits             &            Obtained local reward=16 bits           &    Obtained local reward=16 bits                            \\ \hline
	\end{tabular}
}
\end{table*}
\section{Conclusion}\label{Conclusion}
In this paper, we proposed a joint radio and NFV resource allocation to maximize EE (EE) by satisfying 
 E2E  
 QoS for different service types. In this regard, we formulated an optimization problem in which power and spectrum resources are allocated in the radio part. In the core part, the chaining, placement, and scheduling of VNFS are performed to guarantee the QoS of all users.   We modeled this joint optimization problem as a Markov decision process (MDP) by considering time-varying characteristics of the available resources and wireless channels. Thus, we applied a soft actor-critic deep reinforcement learning (SAC-DRL) algorithm based on the maximum entropy framework to solve the proposed MDP problem.
Simulation results reveled that the proposed joint approach based on SAC-DRL algorithm  could significantly reduce energy consumption compared to the case in which R-RA and NFV-RA problems are optimized separately. 
\bibliography{citation_E2E-RA}	

\begin{thebibliography}{10}
\providecommand{\url}[1]{#1}
\csname url@samestyle\endcsname
\providecommand{\newblock}{\relax}
\providecommand{\bibinfo}[2]{#2}
\providecommand{\BIBentrySTDinterwordspacing}{\spaceskip=0pt\relax}
\providecommand{\BIBentryALTinterwordstretchfactor}{4}
\providecommand{\BIBentryALTinterwordspacing}{\spaceskip=\fontdimen2\font plus
\BIBentryALTinterwordstretchfactor\fontdimen3\font minus
  \fontdimen4\font\relax}
\providecommand{\BIBforeignlanguage}[2]{{%
\expandafter\ifx\csname l@#1\endcsname\relax
\typeout{** WARNING: IEEEtran.bst: No hyphenation pattern has been}%
\typeout{** loaded for the language `#1'. Using the pattern for}%
\typeout{** the default language instead.}%
\else
\language=\csname l@#1\endcsname
\fi
#2}}
\providecommand{\BIBdecl}{\relax}
\BIBdecl

\bibitem{8320765}
I.~{Afolabi}, T.~{Taleb}, K.~{Samdanis}, A.~{Ksentini}, and H.~{Flinck},
  ``Network slicing and softwarization: A survey on principles, enabling
  technologies, and solutions,'' \emph{IEEE Communications Surveys Tutorials},
  vol.~20, no.~3, pp. 2429--2453, Mar. 2018.

\bibitem{8125672}
Z.~{Chang}, Z.~{Zhou}, S.~{Zhou}, T.~{Chen}, and T.~{Ristaniemi}, ``Towards
  service-oriented {5G}: Virtualizing the networks for
  everything-as-a-service,'' \emph{IEEE Access}, vol.~6, pp. 1480--1489, Dec.
  2018.

\bibitem{mijumbi2015design}
R.~{Mijumbi}, J.~{Serrat}, J.~{Gorricho}, N.~{Bouten}, F.~{De Turck}, and
  S.~{Davy}, ``Design and evaluation of algorithms for mapping and scheduling
  of virtual network functions,'' in \emph{Proc. IEEE Conference on Network
  Softwarization (NetSoft)}, Apr. 2015, pp. 1--9.

\bibitem{mijumbi2016network}
R.~{Mijumbi}, J.~{Serrat}, J.~{Gorricho}, N.~{Bouten}, F.~{De Turck}, and
  R.~{Boutaba}, ``Network function virtualization: {State}-of-the-art and
  research challenges,'' \emph{IEEE Communications Surveys Tutorials}, vol.~18,
  no.~1, pp. 236--262, Sep. 2016.

\bibitem{herrera2016resource}
J.~{Gil Herrera} and J.~F. {Botero}, ``Resource allocation in{ NFV: A}
  comprehensive survey,'' \emph{IEEE Transactions on Network and Service
  Management}, vol.~13, no.~3, pp. 518--532, Sep. 2016.

\bibitem{8675284}
A.~N. {Al-Quzweeni}, A.~Q. {Lawey}, T.~E.~H. {Elgorashi}, and J.~M.~H.
  {Elmirghani}, ``Optimized energy aware {5G} network function
  virtualization,'' \emph{IEEE Access}, vol.~7, pp. 44\,939--44\,958, Mar.
  2019.

\bibitem{riera2014virtual}
J.~F. {Riera}, E.~{Escalona}, J.~{Batallé}, E.~{Grasa}, and J.~A.
  {García-Espín}, ``Virtual network function scheduling: {Concept} and
  challenges,'' in \emph{Proc. 2014 International Conference on Smart
  Communications in Network Technologies (SaCoNeT)}, June 2014, pp. 1--5.

\bibitem{alliance2016description}
N.~Alliance, ``Description of network slicing concept,'' \emph{NGMN 5G P},
  vol.~1, p.~1, Jan. 2016.

\bibitem{7926921}
J.~{Ordonez-Lucena}, P.~{Ameigeiras}, D.~{Lopez}, J.~J. {Ramos-Munoz},
  J.~{Lorca}, and J.~{Folgueira}, ``Network slicing for 5{G} with {SDN}/{NFV}:
  Concepts, architectures, and challenges,'' \emph{IEEE Communications
  Magazine}, vol.~55, no.~5, pp. 80--87, May. 2017.

\bibitem{7243304}
R.~{Mijumbi}, J.~{Serrat}, J.~{Gorricho}, N.~{Bouten}, F.~{De Turck}, and
  R.~{Boutaba}, ``Network function virtualization: State-of-the-art and
  research challenges,'' \emph{IEEE Communications Surveys Tutorials}, vol.~18,
  no.~1, pp. 236--262, Sep. 2016.

\bibitem{7945848}
M.~{Mechtri}, C.~{Ghribi}, O.~{Soualah}, and D.~{Zeghlache}, ``{NFV}
  orchestration framework addressing {SFC} challenges,'' \emph{IEEE
  Communications Magazine}, vol.~55, no.~6, pp. 16--23, Jun. 2017.

\bibitem{hossain20155g}
E.~{Hossain} and M.~{Hasan}, ``{5G} cellular: key enabling technologies and
  research challenges,'' \emph{IEEE Instrumentation Measurement Magazine},
  vol.~18, no.~3, pp. 11--21, June 2015.

\bibitem{7143328}
H.~{Dahrouj}, A.~{Douik}, O.~{Dhifallah}, T.~Y. {Al-Naffouri}, and
  M.~{Alouini}, ``Resource allocation in heterogeneous cloud radio access
  networks: advances and challenges,'' \emph{IEEE Wireless Communications},
  vol.~22, no.~3, pp. 66--73, June 2015.

\bibitem{8715830}
P.~{Chemouil}, P.~{Hui}, W.~{Kellerer}, Y.~{Li}, R.~{Stadler}, D.~{Tao},
  Y.~{Wen}, and Y.~{Zhang}, ``Special issue on artificial intelligence and
  machine learning for networking and communications,'' \emph{IEEE Journal on
  Selected Areas in Communications}, vol.~37, no.~6, pp. 1185--1191, 2019.

\bibitem{8932445}
J.~{Pei}, P.~{Hong}, M.~{Pan}, J.~{Liu}, and J.~{Zhou}, ``Optimal {VNF}
  placement via deep reinforcement learning in {SDN/NFV}-enabled networks,''
  \emph{IEEE Journal on Selected Areas in Communications}, vol.~38, no.~2, pp.
  263--278, 2020.

\bibitem{8354944}
A.~{Martin}, J.~{Egaña}, J.~{Flórez}, J.~{Montalbán}, I.~G. {Olaizola},
  M.~{Quartulli}, R.~{Viola}, and M.~{Zorrilla}, ``Network resource allocation
  system for {QoE}-aware delivery of media services in {5G} networks,''
  \emph{IEEE Transactions on Broadcasting}, vol.~64, no.~2, pp. 561--574, 2018.

\bibitem{xie2016service}
Y.~Xie, Z.~Liu, S.~Wang, and Y.~Wang, ``Service function chaining resource
  allocation: A survey,'' \emph{arXiv preprint arXiv:1608.00095}, 2016.

\bibitem{8480442}
M.~M. {Tajiki}, S.~{Salsano}, L.~{Chiaraviglio}, M.~{Shojafar}, and
  B.~{Akbari}, ``Joint energy efficient and {QoS}-aware path allocation and
  {VNF} placement for service function chaining,'' \emph{IEEE Transactions on
  Network and Service Management}, vol.~16, no.~1, pp. 374--388, 2019.

\bibitem{7417401}
M.~T. {Beck} and J.~F. {Botero}, ``Coordinated allocation of service function
  chains,'' in \emph{Proc IEEE Global Communications Conference (GLOBECOM)},
  San Diego, CA, USA, Dec. 2015, pp. 1--6.

\bibitem{8170213}
H.~{Li}, L.~{Wang}, X.~{Wen}, Z.~{Lu}, and L.~{Ma}, ``Constructing service
  function chain test database: An optimal modeling approach for coordinated
  resource allocation,'' \emph{IEEE Access}, vol.~6, pp. 17\,595--17\,605,
  2018.

\bibitem{liu2017dynamic}
J.~{Liu}, W.~{Lu}, F.~{Zhou}, P.~{Lu}, and Z.~{Zhu}, ``On dynamic service
  function chain deployment and readjustment,'' \emph{IEEE Transactions on
  Network and Service Management}, vol.~14, no.~3, pp. 543--553, Sep. 2017.

\bibitem{81702199999}
J.~{Pei}, P.~{Hong}, K.~{Xue}, D.~{Li}, D.~S.~L. {Wei}, and F.~{Wu},
  ``Two-phase virtual network function selection and chaining algorithm based
  on deep learning in {SDN/NFV}-enabled networks,'' \emph{IEEE Journal on
  Selected Areas in Communications}, vol.~38, no.~6, pp. 1102--1117, 2020.

\bibitem{riggio2016scheduling}
R.~{Riggio}, A.~{Bradai}, D.~{Harutyunyan}, T.~{Rasheed}, and T.~{Ahmed},
  ``Scheduling wireless virtual networks functions,'' \emph{IEEE Transactions
  on Network and Service Management}, vol.~13, no.~2, pp. 240--252, June 2016.

\bibitem{cohen2015near}
R.~{Cohen}, L.~{Lewin-Eytan}, J.~S. {Naor}, and D.~{Raz}, ``Near optimal
  placement of virtual network functions,'' in \emph{Proc. IEEE Conference on
  Computer Communications (INFOCOM)}, 2015, pp. 1346--1354. Kowloon, Hong Kong.
  Apr.

\bibitem{8611305}
S.~{Agarwal}, F.~{Malandrino}, C.~F. {Chiasserini}, and S.~{De}, ``{VNF}
  placement and resource allocation for the support of vertical services in
  {5G} networks,'' \emph{IEEE/ACM Transactions on Networking}, vol.~27, no.~1,
  pp. 433--446, 2019.

\bibitem{8255993}
A.~{Alleg}, T.~{Ahmed}, M.~{Mosbah}, R.~{Riggio}, and R.~{Boutaba},
  ``Delay-aware {VNF} placement and chaining based on a flexible resource
  allocation approach,'' in \emph{2017 13th International Conference on Network
  and Service Management (CNSM)}, 2017, pp. 1--7.

\bibitem{cao2017vnf}
J.~Cao, Y.~Zhang, W.~An, X.~Chen, J.~Sun, and Y.~Han, ``{VNF-FG} design and
  {VNF} placement for {5G} mobile networks,'' \emph{Science China Information
  Sciences}, vol.~60, no.~4, p. 040302, 2017.

\bibitem{8316876}
M.~{Dieye}, S.~{Ahvar}, J.~{Sahoo}, E.~{Ahvar}, R.~{Glitho}, H.~{Elbiaze}, and
  N.~{Crespi}, ``{CPVNF}: Cost-efficient proactive {VNF} placement and chaining
  for value-added services in content delivery networks,'' \emph{IEEE
  Transactions on Network and Service Management}, vol.~15, no.~2, pp.
  774--786, 2018.

\bibitem{8424410}
L.~{Tang}, H.~{Yang}, R.~{Ma}, L.~{Hu}, W.~{Wang}, and Q.~{Chen}, ``Queue-aware
  dynamic placement of virtual network functions in {5G} access network,''
  \emph{IEEE Access}, vol.~6, pp. 44\,291--44\,305, 2018.

\bibitem{8281644}
D.~{Li}, P.~{Hong}, K.~{Xue}, and j.~{Pei}, ``Virtual network function
  placement considering resource optimization and {SFC} requests in cloud
  datacenter,'' \emph{IEEE Transactions on Parallel and Distributed Systems},
  vol.~29, no.~7, pp. 1664--1677, July 2018.

\bibitem{8501940}
X.~{Chen}, W.~{Ni}, I.~B. {Collings}, X.~{Wang}, and S.~{Xu}, ``Automated
  function placement and online optimization of network functions
  virtualization,'' \emph{IEEE Transactions on Communications}, vol.~67, no.~2,
  pp. 1225--1237, Feb. 2019.

\bibitem{7859379}
C.~{Pham}, N.~H. {Tran}, S.~{Ren}, W.~{Saad}, and C.~S. {Hong}, ``Traffic-aware
  and energy-efficient {VNF} placement for service chaining: Joint sampling and
  matching approach,'' \emph{IEEE Transactions on Services Computing}, pp.
  1--1, 2017.

\bibitem{nejad2018vspace}
M.~A.~T. Nejad, S.~Parsaeefard, M.~A. Maddah-Ali, T.~Mahmoodi, and B.~H.
  Khalaj, ``{vSPACE}: {VNF} simultaneous placement, admission control and
  embedding,'' \emph{IEEE Journal on Selected Areas in Communications},
  vol.~36, no.~3, pp. 542--557, Mar. 2018.

\bibitem{9000731}
J.~{Chen}, H.~{Liu}, and H.~{Jia}, ``Cross-layer resource allocation in
  wireless-enabled {NFV},'' \emph{IEEE Wireless Communications Letters}, pp.
  1--1, IEEE Early Access, Feb. 2020.

\bibitem{7490404}
L.~{Qu}, C.~{Assi}, and K.~{Shaban}, ``Delay-aware scheduling and resource
  optimization with network function virtualization,'' \emph{IEEE Transactions
  on Communications}, vol.~64, no.~9, pp. 3746--3758, Sep. 2016.

\bibitem{game-theroy}
C.~{Pham}, N.~H. {Tran}, and C.~S. {Hong}, ``Virtual network function
  scheduling: A matching game approach,'' \emph{IEEE Communications Letters},
  vol.~22, no.~1, Jan. 2018.

\bibitem{8256017}
H.~A. {Alameddine}, L.~{Qu}, and C.~{Assi}, ``Scheduling service function
  chains for ultra-low latency network services,'' in \emph{2017 13th
  International Conference on Network and Service Management (CNSM)}, 2017, pp.
  1--9.

\bibitem{kim2018performance}
H.~Kim, ``Performance evaluation of revised virtual resources allocation scheme
  in network function virtualization ({NFV}) networks,'' \emph{Cluster
  Computing}, vol.~22, no.~1, pp. 2331--2339, 2019.

\bibitem{8951149}
S.~{Khan Tayyaba}, H.~A. {Khattak}, A.~{Almogren}, M.~A. {Shah}, I.~{Ud Din},
  I.~{Alkhalifa}, and M.~{Guizani}, ``{5G} vehicular network resource
  management for improving radio access through machine learning,'' \emph{IEEE
  Access}, vol.~8, pp. 6792--6800, 2020.

\bibitem{9128963}
J.~{Chen}, J.~{Chen}, R.~{Hu}, and H.~{Zhang}, ``{QMORA}: A {Q-Learning} based
  multi-objective resource allocation scheme for {NFV} orchestration,'' in
  \emph{2020 IEEE 91st Vehicular Technology Conference (VTC2020-Spring)}, 2020,
  pp. 1--6.

\bibitem{8855889}
J.~{Li}, W.~{Shi}, N.~{Zhang}, and X.~S. {Shen}, ``Reinforcement learning based
  {VNF} scheduling with {End-to-End} delay guarantee,'' in \emph{2019 IEEE/CIC
  International Conference on Communications in China (ICCC)}, 2019, pp.
  572--577.

\bibitem{8901169}
S.~{Troia}, R.~{Alvizu}, and G.~{Maier}, ``Reinforcement learning for service
  function chain reconfiguration in {NFV-SDN} metro-core optical networks,''
  \emph{IEEE Access}, vol.~7, pp. 167\,944--167\,957, 2019.

\bibitem{SOFT}
T.~Haarnoja, A.~Zhou, P.~Abbeel, and S.~Levine, ``Soft actor-critic: Off-policy
  maximum entropy deep reinforcement learning with a stochastic actor,'' in
  \emph{International Conference on Machine Learning}.\hskip 1em plus 0.5em
  minus 0.4em\relax PMLR, 2018, pp. 1861--1870.

\bibitem{7949048}
N.~{Yu}, Z.~{Song}, H.~{Du}, H.~{Huang}, and X.~{Jia}, ``Dynamic resource
  provisioning for energy efficient cloud radio access networks,'' \emph{IEEE
  Transactions on Cloud Computing}, vol.~7, no.~4, pp. 964--974, Oct. 2019.

\bibitem{ETSIG003}
{ETSI, GSNFV}, ``Network functions virtualisation ({NFV}); terminology for main
  concepts in {NFV},'' \emph{ETSI GS NFV}, vol.~2, no.~2, p.~V1, Aug. 2018.

\bibitem{etsi2013network}
------, ``Network functions virtualisation ({NFV}): Architectural framework,''
  \emph{ETSI GS NFV}, vol.~2, no.~2, p.~V1, 2018.

\bibitem{yoon2016nfv}
M.~S. Yoon and A.~E. Kamal, ``{NFV} resource allocation using mixed queuing
  network model,'' in \emph{in Proc. IEEE Global Communications Conference
  (GLOBECOM)}.\hskip 1em plus 0.5em minus 0.4em\relax IEEE, 2016, pp.
  Washington, DC, USA, 1--6. Dec. 2016.

\bibitem{9076109}
Z.~{Shu} and T.~{Taleb}, ``A novel {QoS} framework for network slicing in {5G}
  and beyond networks based on {SDN} and {NFV},'' \emph{IEEE Network}, vol.~34,
  no.~3, pp. 256--263, 2020.

\bibitem{8423711}
G.~{Sun}, Y.~{Li}, D.~{Liao}, and V.~{Chang}, ``Service function chain
  orchestration across multiple domains: A full mesh aggregation approach,''
  \emph{IEEE Transactions on Network and Service Management}, vol.~15, no.~3,
  pp. 1175--1191, 2018.

\bibitem{7387398}
S.~{Van Rossem}, W.~{Tavernier}, B.~{Sonkoly}, D.~{Colle}, J.~{Czentye},
  M.~{Pickavet}, and P.~{Demeester}, ``Deploying elastic routing capability in
  an {SDN/NFV}-enabled environment,'' in \emph{2015 {IEEE} Conference on
  Network Function Virtualization and Software Defined Network ({NFV-SDN})},
  2015, pp. 22--24.

\bibitem{7279063}
M.~{Dayarathna}, Y.~{Wen}, and R.~{Fan}, ``Data center energy consumption
  modeling: A survey,'' \emph{IEEE Communications Surveys Tutorials}, vol.~18,
  no.~1, pp. 732--794, 2016.

\bibitem{17reinforcement}
T.~Haarnoja, H.~Tang, P.~Abbeel, and S.~Levine, ``Reinforcement learning with
  deep energy-based policies,'' in \emph{International Conference on Machine
  Learning}.\hskip 1em plus 0.5em minus 0.4em\relax PMLR, 2017, pp. 1352--1361.

\bibitem{conventional}
C.~Zhong, Z.~Lu, M.~C. Gursoy, and S.~Velipasalar, ``A deep actor-critic
  reinforcement learning framework for dynamic multichannel access,''
  \emph{IEEE Transactions on Cognitive Communications and Networking}, vol.~5,
  no.~4, pp. 1125--1139, 2019.

\bibitem{sutton2018reinforcement}
R.~S. Sutton and A.~G. Barto, \emph{Reinforcement learning: An
  introduction}.\hskip 1em plus 0.5em minus 0.4em\relax MIT press, 2018.

\bibitem{SoAcCrDRL}
F.~Fu, Y.~Kang, Z.~Zhang, F.~R. Yu, and T.~Wu, ``Soft actor–critic {DRL} for
  live transcoding and streaming in vehicular {Fog}-computing-enabled {IoV},''
  \emph{IEEE Internet of Things Journal}, vol.~8, no.~3, pp. 1308--1321, 2021.

\bibitem{817021999}
Z.~{Li} and C.~{Guo}, ``Multi-agent deep reinforcement learning based spectrum
  allocation for {D2D} underlay communications,'' \emph{IEEE Transactions on
  Vehicular Technology}, vol.~69, no.~2, pp. 1828--1840, 2020.

\bibitem{grondman2012survey}
I.~Grondman, L.~Busoniu, G.~A. Lopes, and R.~Babuska, ``A survey of
  actor-critic reinforcement learning: Standard and natural policy gradients,''
  \emph{IEEE Transactions on Systems, Man, and Cybernetics, Part C
  (Applications and Reviews)}, vol.~42, no.~6, pp. 1291--1307, 2012.

\bibitem{817021888}
C.~J. {Watkins} and P.~{Dayan}, ``Q-learning,'' \emph{Machine Learning},
  vol.~8, no.~4, p. 279–292, 1992.

\bibitem{8170218888}
J.~{Hu}, H.~{Zhang}, L.~{Song}, R.~{Schober}, and H.~V. {Poor}, \emph{IEEE
  Transactions on Communications}.

\bibitem{861130000}
M.~{Savi}, M.~{Tornatore}, and G.~{Verticale}, ``Impact of processing-resource
  sharing on the placement of chained virtual network functions,'' \emph{IEEE
  Transactions on Cloud Computing}, 2019.

\bibitem{ebrahimi2020joint}
S.~Ebrahimi, A.~Zakeri, B.~Akbari, and N.~Mokari, ``Joint resource and
  admission management for slice-enabled networks,'' in \emph{NOMS 2020-2020
  IEEE/IFIP Network Operations and Management Symposium}.\hskip 1em plus 0.5em
  minus 0.4em\relax IEEE, 2020, pp. 1--7.

\bibitem{xiao2019nfvdeep}
Y.~Xiao, Q.~Zhang, F.~Liu, J.~Wang, M.~Zhao, Z.~Zhang, and J.~Zhang, ``Nfvdeep:
  Adaptive online service function chain deployment with deep reinforcement
  learning,'' in \emph{Proceedings of the International Symposium on Quality of
  Service}, 2019, pp. 1--10.

\bibitem{ng2011resource}
D.~W.~K. Ng and R.~Schober, ``Resource allocation and scheduling in multi-cell
  ofdma systems with decode-and-forward relaying,'' \emph{IEEE Transactions on
  Wireless Communications}, vol.~10, no.~7, pp. 2246--2258, 2011.

\end{thebibliography}
\bibliographystyle{ieeetran}
\end{document}